\documentclass{article}
\usepackage{latexsym}

\def\dag{\dagger} \def\pd{\partial} \def\pp{\prime} \def\a{\alpha} \def\b{\beta} \def\dl{\delta} \def\s{\sigma} \def\vphi{\varphi} \def\eps{\epsilon} \def\veps{\varepsilon} \def\lam{\lambda} \def\Lam{\Lambda} \def\gm{\gamma} \def\Gm{\Gamma} \def\om{\omega} \def\Om{\Omega} \def\sq{\sqrt} \def\fr{\frac} \def\half{\frac{1}{2}}
 
\def\hg{{\hat g}} \def\bg{{\bar g}} \def\hgm{{\hat \gamma}} \def\nb{\nabla} \def\hnb{{\hat \nabla}} \def\bnb{{\bar \nabla}} \def\hDelta{{\hat \Delta}} \def\bDelta{{\bar \Delta}} \def\hR{{\hat R}} \def\bR{{\bar R}} \def\hE{{\hat E}} \def\bE{{\bar E}}  \def\bC{{\bar C}} \def\C{{\bf C}} \def\D{{\bf D}} \def\E{{\bf E}} \def\G{{\bf G}} \def\H{{\bf H}} \def\M{{\rm M}} \def\T{{\rm T}} \def\S{{\rm S}} \def\L{{\rm L}} \def\V3{{\rm V}_3} \def\prm{m^\prime}  \def\bx{{\bf x}} \def\by{{\bf y}} \def\QG{{\rm QG}}  \def\pl{{\rm pl}}

\begin{document}

\begin{titlepage}

\begin{flushright}
November 2008
\end{flushright}

\vspace{5mm}

\begin{center}
{\Large {\bf Conformal Field Theory on $R \times S^3$ \\ from Quantized Gravity}} 
\end{center}

\vspace{5mm}

\begin{center}
{\sc Ken-ji Hamada}\footnote{E-mail address: hamada@post.kek.jp}
\end{center}

\begin{center}
{\it Institute of Particle and Nuclear Studies, KEK, Tsukuba 305-0801, Japan} \\ and \\
{\it Department of Particle and Nuclear Physics, The Graduate University for Advanced Studies (Sokendai), Tsukuba 305-0801, Japan}
\end{center}

\begin{abstract}
Conformal algebra on $R \times S^3$ derived from quantized gravitational fields is examined. The model we study is a renormalizable quantum theory of gravity in four dimensions described by a combined system of the Weyl action for the traceless tensor mode and the induced Wess-Zumino action managing non-perturbative dynamics of the conformal factor in the metric field. It is shown that the residual diffeomorphism invariance in the radiation$^+$ gauge is equal to the conformal symmetry, and the conformal transformation preserving the gauge-fixing condition that forms a closed algebra quantum mechanically is given by a combination of naive conformal transformation and a certain field-dependent gauge transformation. The unitarity issue of gravity is discussed in the context of conformal field theory. We construct physical states by solving the conformal invariance condition and calculate their scaling dimensions. It is shown that the conformal symmetry mixes the positive-metric and the negative-metric modes and thus the negative-metric mode does not appear independently as a gauge invariant state at all. 
\end{abstract}

\end{titlepage}


\section{Introduction}
\setcounter{equation}{0}
\noindent

Applications of conformal field theories to physics and mathematics in the areas of statistical mechanics, gauge theories and quantum gravity become increasingly indispensable to understand their non-perturbative properties. Especially, conformal invariance in quantum gravity is realized as an exact symmetry of diffeomorphism invariance. Thus, conformal symmetry is an essential tool to understand quantum dynamics of space-time beyond the Planck energy scale.

In two dimensions, the conformal algebra, called the Virasoro algebra, becomes infinite dimensional, leading to significant restrictions on two dimensional conformally invariant theories \cite{isz, bz}. In two-dimensional quantum gravity \cite{polyakov, kpz}, described by the Liouville theory in the conformal gauge \cite{dk, seiberg, gl, lz, bmp, hamada94}, the generator of diffeomorphism symmetry forms the Virasoro algebra without central charge such as $[ L_n, L_m ] = (n-m)L_{n+m}$. Diffeomorphism invariant physical states are classified in terms of the conformally invariant physical state satisfying the conditions $L_n |{\rm phys} \rangle =0 ~ (n \geq 0)$ \cite{lz, bmp}. We can do the similar analysis in four-dimensional quantum gravity \cite{riegert, am, amm92, amm97, hs, hamada02, hh, hamada05, nova}.

In four dimensions, the conformal algebra becomes finite dimensional, while the isometry of the base manifold forms a stringent non-Abelian group instead. Therefore, conformal symmetry in four dimensions also leads to significant restrictions on conformal field theories.

The conformal field theory we study here is that obtained as the core part of renormalizable quantum theory of gravity formulated in a partially non-perturbative manner that the conformal mode in the metric field is treated exactly without introducing its own coupling constant, while the traceless tensor mode is handled perturbatively in terms of a dimensionless coupling constant. It describes quantum states of space-time in a non-perturbative regime beyond the Planck scale. In this paper we examine this conformal field theory focusing on the equivalence between conformal symmetry and diffeomorphism invariance, and discuss the physical properties of diffeomorphism invariant quantum states in the context of conformal field theory.

This paper is organized as follows: after the brief summary on renormalizable quantum gravity is given in the next section, we examine diffeomorphism invariance at the vanishing limit of the coupling constant and show that conformal symmetry is equal to diffeomorphism invariance in section 3. The section 4 devotes to present the canonical quantization of gravitational fields on $R\times S^3$ following the Dirac's quantization procedure. In section 5 we construct the generators of conformal symmetry that form the closed algebra of $SO(4,2)$. These generators yield conformal transformations preserving the gauge-fixing condition that consist of a combination of naive conformal transformations and field-dependent gauge transformations, as shown by Fradkin and Palchik \cite{fp, kluwer}. Then, it is shown that in order that the conformal algebra representing diffeomorphism invariance closes quantum mechanically, the higher-derivative gravitational field is necessary. Conformally invariant physical states in quantum gravity are equal to diffeomorphism invariant quantum states. The some of them are summarized in section 6, and then we discuss scaling properties of these conformal states and their reality required for unitarity. The last section devotes to conclusion and discussion. We here give a comment on the early unitarity arguments for higher-derivative models developed in 1970's.

In this paper the signature of the metric is taken as $(-1,1,1,1)$, and the curvature conventions are $R_{\mu\nu}=R^\lam_{~\mu\lam\nu}$ and $R^\lam_{~\mu\s\nu}=\pd_\s \Gm^\lam_{\mu\nu}-\cdots$.

\section{The Model}
\setcounter{equation}{0}
\noindent

If we wish to apply the Einstein theory for the Planck scale phenomena \cite{dewitt,tv}, it has fatal difficulties such as the black-hole singularity and divergences in the canonical quantization procedure. 
Historically, since it has been recognized that any attempt to quantize Einstein gravity perturbatively cannot be succeeded, many authors \cite{stelle, tomboulis, ft, ft-report} tackled the divergence problem introducing four-derivative terms in the action of gravity, because the gravitational coupling constant becomes dimensionless, and at the same time we can avoid the unbounded problem of the action. However, the $R^2$ action with correct sign to make the action bounded below\footnote{ 
In the Wick-rotated Euclidean action, it is simply denoted that the path integral has the correct weight $e^{-I_{\rm E}}$ with the action $I_{\rm E}$ bounded below. 
} 
indicates the asymptotically non-free behavior. Furthermore, the higher derivative actions create indispensable negative-metric modes.

Through studies of two-dimensional quantum gravity \cite{polyakov, kpz, dk}, it has become clear that all of these problems arise simply because the formulation does not correctly take into account the diffeomorphism invariance, or background metric independence, quantum mechanically. To begin with, in an analogy of two-dimensional model, a four-dimensional counter model for the conformal factor given by Riegert \cite{riegert} has been quantized by Antoniadis et al. \cite{am, amm92, amm97}.

In order to construct a four-dimensional quantum gravity realized in the ultraviolet limit, however, we have to manage the dynamics of the traceless tensor mode appropriately \cite{hs, hamada02, nova}. Treating its dynamics in perturbation, we have formulated renormalizable quantum gravity based on the conformal gravity which leads to the full conformal field theory described by a combined system of the Wess-Zumino action and the Weyl action at the vanishing limit of the coupling constant \cite{hh, hamada05}.

\subsection{The action}
\noindent

Quantum gravity is defined by the path integral over gravitational fields with diffeomorphism invariant weight $e^{iI}$. The model we consider here is defined by the dimensionless action \cite{hamada02, nova}:
\begin{eqnarray}
   I = \int d^4 x \sq{-g} \biggl\{
         -\fr{1}{t^2} C^2_{\mu\nu\lam\s} -b G_4 
         + \fr{1}{\hbar} \biggl( \fr{1}{16\pi G}R -\Lam +{\cal L}_\M
          \biggr) \biggr\},
         \label{quantum gravity action}
\end{eqnarray}
where we write the Newton constant as $G$, and the cosmological constant as $\Lam$. The Lagrangian for a matter field action is denoted by ${\cal L}_\M$.

The first two terms are conformally invariant gravitational actions, which are the square of the Weyl tensor
\begin{equation}
  C_{\mu\nu\lam\s}^2 = R_{\mu\nu\lam\s}R^{\mu\nu\lam\s} - 2R_{\mu\nu}R^{\mu\nu} 
                       + \fr{1}{3}R^2 
\end{equation}
and the Euler density
\begin{equation}
    G_4 = R_{\mu\nu\lam\s}R^{\mu\nu\lam\s} -4 R_{\mu\nu}R^{\mu\nu} + R^2 , 
    \label{G_4}
\end{equation}
respectively. The Weyl tensor represents the field strength of traceless tensor modes in gravitational fields, and $t$ is the dimensionless coupling constant. The constant $b$ is introduced to renormalize divergences proportional to $G_4$, which is not an independent coupling constant because it does not have a kinetic term.

The constant $\hbar$ is the Planck constant, which does not appear in front of the four-derivative gravitational actions because, contrary to matter fields, gravitational fields are dimensionless and thus these actions in four dimensions are exactly dimensionless. This implies that the higher-derivative gravitational fields describe purely quantum states, and have no classical meanings. In the following, $\hbar$ is taken to be unity.

The four-derivative gravitational actions are determined by the integrability condition for conformal anomalies \cite{bcr, riegert}. Consider a generic local-form of conformal anomaly \cite{cd, ddi, duff} given by the Weyl transformation of the effective action as
\begin{equation}
    \dl_\om \Gm = \int d^4 x \sq{-g} ~\om \Bigl\{ 
        \eta_1 R_{\mu\nu\lam\s}R^{\mu\nu\lam\s} +\eta_2 R_{\mu\nu}R^{\mu\nu} + \eta_3 R^2 
        +\eta_4 \nb^2 \! R \Bigr\} ,
              \label{variation of effective action}
\end{equation}
where the Weyl transformation is defined by $\dl_\om g_{\mu\nu}=2\om g_{\mu\nu}$. Since the conformal anomalies arise following ultraviolet divergences, it is a possible candidate for the counterterm to renormalize divergences, or the bare action. The integrability condition is defined such that two independent Weyl transformations commute as
\begin{equation}
   [\dl_{\om_1}, \dl_{\om_2}] \Gm  = 8 ( \eta_1 +  \eta_2 + 3\eta_3 ) \times
     \int d^4 x \sq{-g} R \om_{[1} \nb^2 \om_{2]} = 0,
\end{equation}
where the anti-symmetric product is denoted as $a_{[\mu}b_{\nu]}=(a_\mu b_\nu -a_\nu b_\mu)/2$. Thus, the integrability gives a constraint on the form of the bare action. This condition indicates the renormalizability such that the effective action exists.

The integrable quantities are just the square of the Weyl tensor and the Euler density, apart from the trivial term with the parameter $\eta_4$.\footnote{ 
This implies that the local and finite $R^2$ term may appear in the effective action, because it is obtained by integrating the $\eta_4$ term with respect to the conformal mode. However, since it is at the higher order of the coupling constant $t$, we here disregard it.
} 
The lower derivative actions such as the Einstein term and the cosmological constant are trivially integrable in this sense. The relationship between the effective action and the Wess-Zumino action for conformal anomaly satisfying the Wess-Zumino consistency condition \cite{wz} is discussed when they are defined.

In this way, the integrability condition reduces a part of ambiguities in four-dimensional gravitational actions, and thus excludes the purely $R^2$ bare action, which is commonly introduced as the kinetic term of the conformal mode. The kinetic term of the conformal mode as well as its interaction terms with four derivatives are, as mentioned below, given by the Wess-Zumino action induced from the path-integral measure. The integrability condition requires that there should be no divergences proportional to $R^2$. The Hathrell problem \cite{hathrell} on this matter indicating the appearance of the $R^2$ divergences at the three-loop level have been resolved in \cite{hamada02} in terms of dimensional regularization, applying the integrability condition generalized in $D$ dimensions in order to determine the bare action.\footnote{ 
Dimensional regularization is a manifestly diffeomorphism invariant regularization at all orders, in which the Wess-Zumino action appears as a coefficient of the series obtained by expanding the bare action with respect to an infinitesimal parameter $4-D$. 
} 

In this paper, we also consider a scalar field conformally coupled to gravity, $X$, and a $U(1)$ gauge field, $A_\mu$, which are defined by the action
\begin{equation}
       I_\M = \int d^4x \sq{-g} \left\{ 
        -\fr{1}{2} \left( g^{\mu\nu}\pd_\mu X \pd_\mu X +\fr{1}{6}R X^2 \right)
               -\fr{1}{4} g^{\mu\lam}g^{\nu\s}F_{\mu\nu}F_{\lam\s} \right\}, 
\end{equation}
where $F_{\mu\nu}=\nb_\mu A_\nu -\nb_\nu A_\mu=\pd_\mu A_\nu -\pd_\nu A_\mu$ is the gauge field strength.

\subsection{Quantization technique}
\noindent

The perturbation theory is defined by an expansion in $t$ about a conformally flat configuration satisfying $C_{\mu\nu\lam\s}=0$. In order to treat such configurations, the metric field is decomposed into the conformal mode $\phi$ and the traceless tensor mode $h^\mu_{~\nu}$ and the background metric $\hg_{\mu\nu}$ as
\begin{equation}
    g_{\mu\nu}= e^{2\phi}\bg_{\mu\nu}
     \label{metric decomposition}
\end{equation}
and
\begin{equation}
    \bg_{\mu\nu}=\bigl( \hg e^{th} \bigr)_{\mu\nu}
     =\hg_{\mu\lam} \left( \dl^\lam_{~\nu} +th^\lam_{~\nu} 
         +\fr{t^2}{2} ( h^2 )^\lam_{~\nu} +\cdots \right),
        \label{traceless metric field}
\end{equation}
where $tr(h)=h^\lam_{~\lam}=0$. The contraction of the indices of $h^\mu_{~\nu}$ is done by using the background metric. In the following, gravitational quantities with the hat and the bar on them are defined in terms of the metric $\hg_{\mu\nu}$ and $\bg_{\mu\nu}$, respectively.

The conformal mode is treated non-perturbatively without introducing the coupling constant for this mode, while the traceless tensor mode is handled perturbatively in terms of the coupling constant $t$. This treatment is justified by the asymptotically free behavior of traceless tensor mode. It implies that the conformal invariance becomes significant at very high energies where the coupling strength becomes small, and the configuration with the vanishing Weyl tensor are chosen such that the singular configuration like a black hole at which the Riemann-Christoffel curvature tensor is divergent is excluded quantum mechanically.

The technique to treat diffeomorphism invariance is the following. We change the path integral measures from the diffeomorphism invariant measures to the practical measures defined on the background metric $\hg_{\mu\nu}$. Consequently, in order to preserve the diffeomorphism invariance, the Wess-Zumino action $S$ is necessary as the Jacobian, and the partition function is expressed as
\begin{eqnarray}
     Z &=& \int \fr{[dg dA dX]_g}{\rm Vol(diff.)} \exp \{ iI(A,X,g) \}
           \nonumber \\
       &=& \int \fr{[d\phi d h dA dX]_{\hg}}{\rm Vol(diff.)}
           \exp \left\{ iS(\phi,\bg)+ iI(A,X,g) \right\}.
\end{eqnarray}

The Wess-Zumino action is induced even at the vanishing limit of the coupling constant. It, denoted by $S_1$, is given by the so-called local Riegert action \cite{riegert} obtained by integrating the conformal anomaly concerning the Euler density with respect to the conformal mode as
\begin{eqnarray}
       S_1 (\phi, \bg) &=& -\fr{b_1}{(4\pi)^2}\int d^4 x 
                         \int^\phi_0 d\phi \sq{-g}E_4
          \nonumber \\
       &=& -\fr{b_1}{(4\pi)^2}\int d^4 x \sq{-\bg}
                       \left( 2\phi \bDelta_4 \phi + \bE_4 \phi \right),
                  \label{4 dim. Wess-Zumino action}
\end{eqnarray}
where the quantity $E_4$ represents the modified Euler density,
\begin{equation}
   E_4=G_4-\fr{2}{3} \nb^2 R,
\end{equation}
which satisfies the relation $\sq{-g}E_4=\sq{-\bg}(4\bDelta_4 \phi +\bE_4)$, and $\sq{-g}\Delta_4$ is the conformally invariant fourth-order operator acting on a scalar field defined by
\begin{equation}
     \Delta_4 = \nb^4 + 2R^{\mu\nu}\nb_\mu \nb_\nu - \fr{2}{3}R \nb^2 
                + \fr{1}{3} \nb^\mu R \nb_\mu .
\end{equation}
This operator satisfies the self-adjoint condition $\int \sq{-g}A \Delta_4 B = \int \sq{-g} (\Delta_4 A)B $ for scalar fields, $A$ and $B$. The coefficient $b_1$ has been computed as \cite{amm92, cd, ddi, duff}
\begin{equation}
     b_1 = \fr{1}{360} \left( N_X + \fr{11}{2}N_W + 62 N_A \right) + \fr{769}{180},
       \label{Wess-Zumino coefficient}
\end{equation}
where $N_X$, $N_W$ and $N_A$ are the numbers of scalar fields, Weyl fermions and gauge fields added to the action, respectively.

The Wess-Zumino action at the lowest order, $S_1$, contains the kinetic term of the conformal mode. Thus, the dynamics of the conformal mode is induced from the diffeomorphism invariant measures. This action is a four-dimensional counter quantity of the so-called Liouville/Polyakov action \cite{polyakov} in two dimensions.\footnote{ 
The Liouville/Polyakov action is given by $S_\L =-(b_\L/4\pi)\int d^2 x \int^\phi_0 d\phi \sq{-g}R = -(b_\L/4\pi) \int d^2 x \sq{-\hg}(\phi \hDelta_2 \phi + \hR \phi )$, where $\Delta_2=-\nb^2$. The coefficient has been computed as $b_\L=(25-c_\M)/6$ for the case of quantum gravity coupled with a conformal field theory with central charge $c_\M$ \cite{polyakov, kpz, dk}.} 

In the following sections, we consider the combined system at the vanishing limit of the coupling constant,
\begin{equation}
    I_{\rm CFT} = S_1(\phi,\hg) + I(X,A,g)|_{t \to 0},
        \label{4DQG action}
\end{equation}
as a classical action defined on the curved space-time with the background metric $\hg_{\mu\nu}$. All terms with mass scales such as the Einstein action are given by renormalizable composite fields with an exponential factor of the conformal mode leading to a power-law behavior of their correlation functions.\footnote{ 
Thus, there is no logarithmic catastrophe at the Planck mass scale. Since the Einstein action can not be considered as an ordinary mass term, we introduce a small fictitious mass $z$ to regularize infrared divergences. This mass term is not diffeomorphism invariant, and thus the infrared divergence appearing in the form of $\log z$ cancels out. 
} 
In terms of conformal field theory, they are  physical conformal fields discussed in section 6. They are considered as potential terms, and we disregard them considering we are in the very high energy regime beyond the Planck scale.

The effective action, which has the manifestly diffeomorphism invariant form written in terms of the full metric $g_{\mu\nu}$, is now obtained by adding quantum corrections to the classical action $S+I$.  For the conformal-mode sector, it is given by the so-called non-local Riegert action as
\begin{eqnarray}
   \Gm &=& S_1(\phi,\hg) 
      -\fr{b_1}{8(4\pi)^2} \int d^4 x \sq{-\hg} \hE_4 \fr{1}{\hDelta_4}\hE_4
            \nonumber \\
       &=& -\fr{b_1}{8(4\pi)^2} \int d^4 x \sq{-g} E_4 \fr{1}{\Delta_4}E_4 
         \label{non-local Riegert action}
\end{eqnarray}
at the lowest order. The non-local term in the first line is the loop correction obtained after carrying out renormalization of the ultraviolet divergence proportional to the Euler density. Since the conformal mode does not have its own coupling constant, this mode is not renormalized whose renormalization factor is unity at all orders \cite{hamada02}.

In this paper we do not discuss another kind of the Wess-Zumino action $\sq{-g} \phi C^2_{\mu\nu\lam\s}$ obtained by integrating the Weyl-squared conformal anomaly with respect to the conformal mode, because it appears at the order of $t_r^2$ and more, following the non-local term $\log (k^2/\mu^2)$ in connection with the beta function $\b_t =-\b_0 t^3_r$ with $\b_0 >0$ in order to preserve the diffeomorphism invariance, such as
\begin{eqnarray}
   \sq{-g}{\cal L}_{\rm eff} &=&  -\left\{ \fr{1}{t_r^2} - 2\b_0 \phi
         + \b_0 \log \left( \fr{k^2}{\mu^2} \right) +\cdots \right\} \sq{-g} C^2_{\mu\nu\lam\s}
                  \nonumber \\
          &=& -\fr{1}{t_r^2(p)} \sq{-g} C^2_{\mu\nu\lam\s}
\end{eqnarray}
in the momentum space, where $k$ is a momentum defined on the background. Disregarding higher order corrections, the running coupling constant is written as $1/t^2_r(p) = \b_0 \log (p^2/\Lam_\QG^2)$, where $p$ is a physical momentum defined by $p=k/e^{\phi}$ and the parameter $\Lam_\QG =\mu \exp \{ -1/2\b_0 t_r^2 \}$ with $\mu$ being a renormalization mass scale is the new dynamical scale. The running coupling constant is a measure of the degree of deviation from  conformal field theory.

The scale parameter $\Lam_\QG$ represents the energy scale where the correlation length becomes short-range and thus the conformal invariance breaks down turning to the classical Einstein phase. We set the ordering of two mass scales as $m_\pl\gg \Lam_\QG$($\simeq 10^{17}$GeV), where $m_\pl=1/\sq{G}$($\simeq 10^{19}$GeV). Then, we obtain an inflationary scenario with a sufficient number of e-foldings driven by quantum gravity effects \cite{hy,hhy,hhsy}.\footnote{
The inflationary model driven by the conformal anomaly was first proposed by Starobinsky in 1979 \cite{starobinsky}.
} 

\section{Conformal Symmetry as Diffeomorphism Invariance}
\setcounter{equation}{0}
\noindent

Let us first clarify the relation between diffeomorhism invariance and conformal invariance. 
Diffeomorphism invariance is defined by the transformation using a contra-variant vector $\xi^\mu$ as
\begin{equation}
  \dl_\xi g_{\mu\nu}=g_{\mu\lam}\nb_\nu \xi^\lam + g_{\nu\lam}\nb_\mu \xi^\lam, 
\end{equation}
for the metric field and
\begin{eqnarray}
    \dl_\xi X &=& \xi^\lam \nb_\lam X, 
        \nonumber  \\
    \dl_\xi A_\mu &=& \xi^\lam \nb_\lam A_\mu + A_\lam \nb_\mu \xi^\lam
\end{eqnarray}
for the scalar field and the covariant vector field.

Under the decomposition (\ref{metric decomposition}), each mode of the metric field transforms as
\begin{eqnarray}
   \dl_\xi \phi &=& \xi^\lam \hnb_\lam \phi  
                       + \fr{1}{4} \hnb_\lam \xi^\lam ,     
            \nonumber \\
   \dl_\xi \bg_{\mu\nu} &=& \bg_{\mu\lam} \bnb_\nu \xi^\lam 
                 +\bg_{\nu\lam} \bnb_\mu \xi^\lam  
                 -\fr{1}{2} \bg_{\mu\nu} \hnb_\lam \xi^\lam ,    
\end{eqnarray}
where we use the equation $\bnb_\lam \xi^\lam = \hnb_\lam \xi^\lam$.  Expanding the second equation in the coupling constant as (\ref{traceless metric field}), we obtain the transformation law of the traceless tensor mode,
\begin{eqnarray}
   \dl_\xi h_{\mu\nu} &=& \fr{1}{t} \left( \hnb_\mu \xi_\nu +\hnb_\nu \xi_\mu
           - \fr{1}{2} \hg_{\mu\nu} \hnb_\lam \xi^\lam \right)
           + \xi^\lam \hnb_\lam h_{\mu\nu}   
                \nonumber    \\ 
    &&     + \fr{1}{2} h_{\mu\lam} \left( \hnb_\nu \xi^\lam 
                  - \hnb^\lam \xi_\nu \right) 
           + \fr{1}{2} h_{\nu\lam} \left( \hnb_\mu \xi^\lam 
                  - \hnb^\lam \xi_\mu \right) 
           + o(t\xi h^2),
               \nonumber \\ 
               \label{diffeomorphism for traceless mode}
\end{eqnarray}
where the covariant vector $\xi_\mu$ is defined using the background metric as $\xi_\mu =\hg_{\mu\nu} \xi^\nu$. Thus, the transformations of the conformal mode and the traceless tensor mode are decoupled.

The conformally invariant Weyl action does not depend on the conformal mode and can be written in terms of the metric field $\bg_{\mu\nu}$ as $(-1/t^2)\int \sq{-\bg}\bC_{\mu\nu\lam\s}^2$, which is invariant under the transformation (\ref{diffeomorphism for traceless mode}).

The gauge field action is also written in terms of the metric field $\bg_{\mu\nu}$ with the covariant vector field $A_\mu$ unchanged, while in order to remove the conformal-mode dependence in the conformally coupled scalar field action we have to rescale the scalar field as $X \to e^{-\phi}X$. Thus, the matter action can be written in terms of the rescaled scalar and the covariant vector fields as
\begin{equation}
       I_\M = \int d^4x \sq{-\bg} \left\{ 
        -\fr{1}{2} \left( \bg^{\mu\nu}\pd_\mu X \pd_\mu X 
                          +\fr{1}{6}\bR X^2 \right)
               -\fr{1}{4} \bg^{\mu\lam}\bg^{\nu\s}F_{\mu\nu}F_{\lam\s} \right\}. 
\end{equation}
Then, the transformation law of diffeomorphism invariance for the scalar field changes to
\begin{equation}
    \dl_\xi X = \xi^\lam \bnb_\lam X + \fr{1}{4} X \bnb_\lam \xi^\lam ,
\end{equation}
compensating contributions from the transformation of the conformal mode. The transformation law of the gauge field becomes 
\begin{equation}
    \dl_\xi A_\mu = \xi^\lam \bnb_\lam A_\mu + A_\lam \bnb_\mu \xi^\lam ,
\end{equation}
and, for the contravariant vector field, $\dl_\xi A^\mu= \dl_\xi (\bg^{\mu\nu} A_\nu)$, where $\dl_\xi \bg^{\mu\nu}=-\bg^{\mu\lam}\bg^{\nu\s}\dl_\xi \bg_{\lam\s}$. In the following, we use this matter action and these transformations as diffeomorphism invariance.

In this section we discuss diffeomorphism invariance at the vanishing limit of the coupling constant. Since the Weyl action is divided by the square of the coupling, only the kinetic term of the traceless tensor mode survives at the limit. The interaction terms with other fields also drop out.

There are two types of diffeomorphism at the vanishing coupling limit. The first is the gauge invariance for the kinetic term of the Weyl action. We introduce the gauge parameter $\kappa^\mu=\xi^\mu/t$ and take the limit $t \to 0$ with leaving $\kappa^\mu$ finite. Then, from the transformation (\ref{diffeomorphism for traceless mode}), the diffeomorphism is expressed as
\begin{equation}
   \dl_\kappa h_{\mu\nu} = \hnb_\mu \kappa_\nu + \hnb_\nu \kappa_\mu 
                          - \fr{1}{2} \hg_{\mu\nu} \hnb_\lam \kappa^\lam, 
            \label{traceless mode gauge transformation}
\end{equation}
while other fields do not transform under the limit as $\dl_\kappa \phi =\dl_\kappa X = \dl_\kappa A_\mu =0$, because the transformations for these fields become of order of $t$ in the expansion using $\kappa^\mu$.

This transformation is similar to the $U(1)$ gauge transformation,
\begin{equation}
   \dl_\lam A_\mu = \hnb_\mu \lam .
     \label{gauge transformation}
\end{equation}
The gauge parameters $\kappa^\mu$ and $\lam$ are used later to fix the gauge degrees of freedom of the traceless tensor mode and the gauge field, respectively.

The second is the conformal invariance we will discuss in this paper. It is the diffeomorphism symmetry with a gauge parameter $\xi^\mu =\zeta^\mu$ satisfying the conformal Killing equation
\begin{equation}
    \hnb_\mu \zeta_\nu + \hnb_\nu \zeta_\mu 
                 - \fr{1}{2} \hg_{\mu\nu} \hnb_\lam \zeta^\lam =0.
           \label{conformal Killing equation}
\end{equation}
Since the lowest term of the transformation of $h_{\mu\nu}$ (\ref{diffeomorphism for traceless mode}) vanishes in this case, the second term becomes effective such that the kinetic term of the Weyl action becomes invariant under the transformation
\begin{equation}
    \dl_\zeta h_{\mu\nu} = \zeta^\lam \hnb_\lam h_{\mu\nu} 
              + \half h_{\mu\lam} \left( \hnb_\nu \zeta^\lam - \hnb^\lam \zeta_\nu \right)
              + \half h_{\nu\lam} \left( \hnb_\mu \zeta^\lam - \hnb^\lam \zeta_\mu \right) 
      \label{traceless mode conformal transformation}
\end{equation}
without taking into account self-interaction terms. And also, the scalar and the gauge fields transform as,
\begin{equation}
   \dl_\zeta X = \zeta^\lam \hnb_\lam X + \fr{1}{4} X \hnb_\lam \zeta^\lam  
     \label{scalar field conformal transformation}
\end{equation}
and
\begin{equation}
   \dl_\zeta A_\mu = \zeta^\lam \hnb_\lam A_\mu + A_\lam \hnb_\mu \zeta^\lam ,
    \label{gauge field conformal transformation}
\end{equation}
respectively. Due to the disappearance of the lowest term in the transformation of the traceless tensor mode, the kinetic term of each field becomes invariant without interaction terms with the traceless tensor mode. Since the background metric does not change, this transformation is a conformal transformation considering quantum gravity as a quantum field theory on the background.

For the case of the scalar field, for example, the conformal invariance can be easily shown in the flat background as
\begin{eqnarray}
    \dl_\zeta I_X &=& -\int d^4x \pd^\mu X \pd_\mu \left( \zeta^\lam \pd_\lam X 
              +\fr{1}{4} X \pd_\lam \zeta^\lam \right) 
               \nonumber \\
        &=& \int d^4 x \biggl\{ 
           -\fr{1}{4}\left( 3\pd_\eta \zeta_0 + \pd_i \zeta^i \right) \pd_\eta X \pd_\eta X
           +\left( \pd_\eta \zeta_i + \pd_i \zeta_0 \right) \pd_\eta X \pd^i X
               \nonumber \\
        && +\left[ -\pd_i \zeta_j + \fr{1}{4}\dl_{ij} \left( -\pd_\eta \zeta_0 
                + \pd_k \zeta^k \right) \right] \pd^i X \pd^j X
           + \fr{1}{8} (\pd_\s \pd^\s  \pd_\lam \zeta^\lam) X^2 
            \biggr\}
               \nonumber \\
        &=& 0 ,
\end{eqnarray}
using the conformal Killing equations.

The conformal mode transforms as
\begin{equation}
   \dl_\zeta \phi = \zeta^\lam \hnb_\lam \phi + \fr{1}{4} \hnb_\lam \zeta^\lam
     \label{conformal mode conformal transformation}
\end{equation}
Since there is a shift term independent of $\phi$, it is not a scalar transformation. The Wess-Zumino action changes under this transformation and produces the quantity
\begin{equation}
      \dl_\zeta S_1 = - \fr{b_1}{(4\pi)^2} \int d^4 x 
                     \hbox{$\sq{-\hg}$}\hE_4 \fr{1}{4}\hnb_\lam \zeta^\lam .
\end{equation}
It is the same form to the conformal anomaly, but the overall sign is opposite to that produced by quantizing fields. These quantities cancel out and thus the conformal invariance, or diffeomorphism invariance recovers quantum mechanically. Thus, adding quantum corrections to the classical Wess-Zumino action (\ref{4DQG action}), we obtain the manifestly diffeomorphism invariant effective action, as shown in (\ref{non-local Riegert action}).

\section{Canonical Quantization on $R \times S^3$}
\setcounter{equation}{0}
\noindent

To quantize the model in practice, we need to specify the background metric. Since the asymptotic freedom implies that the Weyl tensor should vanish at the vanishing limit of the coupling constant, it is specified to be a conformally flat metric. Owing to the conformal invariance, all models transformed by a conformal transformation into each other are equivalent.

We here choose the cylindrical background $R \times S^3$ because it has several advantages. Mode expansions of higher derivative fields become simple and the canonical commutation relations have diagonal forms, contrary to the case of flat background \cite{pu} in which there is an unusual time-dependence in the mode expansion and the commutator in general becomes off-diagonal. Also, we can use tools developed in the $SU(2)$ representation theory \cite{vmk}, because the isometry group of $S^3$ is $SO(4)=SU(2) \times SU(2)$.

The background metric is parametrized using the Euler angles $x^i=(\a,\b,\gm)$ as 
\begin{eqnarray}
     d{\hat s}^2_{R\times S^3} &=& \hg_{\mu\nu}dx^{\mu}dx^{\nu}
                  =-d\eta^2 + \hgm_{ij}dx^i dx^j
            \nonumber  \\
                 &=&-d\eta^2 + \fr{1}{4} (d\a^2 +d\b^2 +d\gm^2 +2 \cos \b d\a d\gm )
           \label{metric}
\end{eqnarray}
The radius of $S^3$ is taken to be unity. The curvatures are then given by $\hR_{0\mu\nu\lam}=\hR_{0\mu}=0$ and
\begin{equation}
   \hR_{ijkl}=(\hgm_{ik}\hgm_{jl}-\hgm_{il}\hgm_{jk}), \quad
   \hR_{ij}=2\hgm_{ij}, \quad
   \hR=6,
\end{equation}
and $\hat{C}^2_{\mu\nu\lam\s}=\hat{G}_4=0$. The volume element on the unit $S^3$ is 
\begin{equation}
      d\Om_3 =d^3 x \hbox{$\sq{\hgm}$} =\fr{1}{8}\sin \b d\a d\b d\gm,
\end{equation}
and the volume is given by
\begin{equation}
      \V3 = \int_{S^3} d\Om_3 =2\pi^2.
\end{equation}

Dynamical fields are expanded in symmetric-traceless-transverse (ST$^2$) spherical tensor harmonics \cite{ro}. The ST$^2$ tensor harmonics of rank $n$ are constructed and classified using $(J+\veps_n,J-\veps_n)$ representation of the isometry group $SU(2) \times SU(2)$ for each sign of the polarization index $\veps_n=\pm n/2$ \cite{hh}. They, denoted by $Y^{i_1 \cdots i_n}_{J(M\veps_n)}$, are the eigenfunction of the Laplacian on $S^3$, $\Box_3 =\hgm^{ij}\hnb_i \hnb_j$, as 
\begin{equation}
         \Box_3 Y^{i_1 \cdots i_n}_{J (M \veps_n)}
      =\{ -2J(2J+2)+n \} Y^{i_1 \cdots i_n}_{J (M \veps_n)},
\end{equation}
where $J~(\geq n/2)$ takes integer or half-integer values, and $M=(m,m^\pp)$ represents the multiplicity for each polarization,
\begin{eqnarray}
  m &=&-J-\veps_n,~ -J-\veps_n+1, \cdots, J+\veps_n-1,~ J+\veps_n,
             \nonumber \\
  m^\pp &=&-J+\veps_n,~ -J+\veps_n+1, \cdots, J-\veps_n-1,~ J-\veps_n
\end{eqnarray}
Thus, the multiplicity is given by $(2J+1)^2$ for $n=0$, and $2(2J+n+1)(2J-n+1)$ for $n \geq 1$ taking into account the polarization.

The complex conjugate of ST$^2$ tensor harmonics and the normalization are defined by
\begin{eqnarray}
    && Y^{i_1 \cdots i_n *}_{J (M \veps_n)}
      = (-1)^n \eps_M Y^{i_1 \cdots i_n}_{J (-M \veps_n)},
            \nonumber \\
    && \int_{S^3} d\Om_3 Y^{i_1 \cdots i_n *}_{J_1 (M_1 \veps^1_n)}
                      Y_{i_1 \cdots i_n J_2 (M_2 \veps^2_n)}
       = \dl_{J_1J_2}\dl_{M_1 M_2}\dl_{\veps^1_n \veps^2_n},
\end{eqnarray}
where the Kronecker delta for the index $M$ is defined by $\dl_{M_1 M_2}=\dl_{m_1 m_2}\dl_{m_1^\pp m_2^\pp}$, and the sign factor is
\begin{equation}
    \eps_M=(-1)^{m-\prm},
\end{equation}
satisfying $\eps_M^2=1$.  In the following, we use the parametrizations,
\begin{equation}
    y=\veps_1=\pm 1/2, \quad x=\veps_2=\pm 1, \quad
    z=\veps_3=\pm 3/2, \quad w=\veps_4=\pm 2 ,
\end{equation}
for the polarization indices for the tensor up to rank $4$.

\subsection{Scalar fields}
\noindent

To begin with, let us consider the canonical quantization of the scalar field. The action on $R \times S^3$ is written by
\begin{equation}
     I_X = \int d\eta \int_{S^3} d\Om_3 \half X \left( -\pd_\eta^2 + \Box_3 -1 \right) X,
\end{equation}
where the missing dimension in the expression originates from the radius taken to be unity.

The scalar field is expanded in scalar harmonics as $X \propto e^{-i\om\eta}Y_{JM}$. Since the equation of motion leads to the dispersion relation $\om^2-(2J+1)^2=0$, we write the scalar field as 
\begin{equation}
      X = \sum_{J \geq 0}\sum_M \fr{1}{\sq{2(2J+1)}} \left\{ 
            \vphi_{JM} e^{-i(2J+1)\eta}Y_{JM}
            + \vphi^\dag_{JM} e^{i(2J+1)\eta}Y^*_{JM} \right\} .
\end{equation}

The canonical quantization is carried out in the standard manner setting the equal-time commutation relation between $X$ and its conjugate momentum as 
\begin{equation}
    [ X(\eta, \bx), P_X(\eta, \by) ]=i\dl_3 (\bx-\by),
\end{equation}
where the momentum variable is defined by $P_X=\dl {\cal L}^X/\dl(\pd_\eta X)=\pd_\eta X$, denoting the Lagrangian on the background manifold as ${\cal L}^X$. The delta function is defined in terms of the complete set of scalar harmonics as
\begin{equation}
     \dl_3 (\bx -\by) = \sum_{J \geq 0} \sum_M Y^*_{JM}(\bx)Y_{JM}(\by).
\end{equation}
The creation and annihilation operators then satisfy the commutation relation
\begin{equation}
    [ \vphi_{J_1 M_1}, \vphi^\dag_{J_2 M_2} ] = \dl_{J_1 J_2}\dl_{M_1 M_2},
\end{equation}
and the Hamiltonian is given by
\begin{eqnarray}
   H^X &=& \int_{S^3} d\Om_3 :\left\{ \half P_X^2 - \half X \left( \Box_3-1 \right) X \right\}:
            \nonumber \\
       &=& \sum_{J \geq 0} \sum_M (2J+1)\vphi^\dag_{JM}\vphi_{JM},
       \label{scalar field Hamiltonian}
\end{eqnarray}
where $:~:$ denotes taking the normal ordering.

\subsection{Gauge fields}
\noindent

In order to quantize gauge fields we have to fix the gauge symmetry. We here take the transverse gauge, called the Coulomb gauge, defined by the condition,
\begin{equation}
     \hnb^i A_i =0.
     \label{Coulomb gauge}
\end{equation}
Then the gauge fixed action on $R \times S^3$ is written as
\begin{equation}
    I_A = \int d\eta \int_{S^3} d\Om_3 \left\{
     \half A^i\left( -\pd_\eta^2 + \Box_3 -2 \right)A_i
     - \half A_0 \Box_3 A_0 \right\},
\end{equation}
where the contravariant vector field is now defined by $A^i=\hgm^{ij} A_j$.

Since the kinetic term of the time-component of the gauge field does not have its time-derivative, it is not a dynamical field. So, we further remove it using the residual gauge degree of freedom preserving the transverse-gauge condition as
\begin{equation}
      A_0=0.
      \label{A_0=0 gauge}
\end{equation}
This is the so-called radiation gauge.

The transverse gauge field is expanded in vector harmonics as $A^i \propto e^{-i\om\eta}Y^i_{J(My)}$. Since the equation of motion leads to the dispersion relation $\om^2 -(2J+1)^2=0$ equal to that of the scalar field, we write the gauge field as
\begin{equation}
    A^i = \sum_{J \geq \half}\sum_{M,y} \fr{1}{\sq{2(2J+1)}} \left\{ 
            q_{J(My)} e^{-i(2J+1)\eta}Y^i_{J(My)}
            + q^\dag_{J(My)} e^{i(2J+1)\eta}Y^{i*}_{J(My)} \right\} .
\end{equation}
The conjugate momentum is given by $P^i_A=\pd_\eta A^i$ and the equal-time commutation relation is set as
\begin{equation}
    [ A^i(\eta, \bx), P^j_A(\eta, \by) ]=i\dl_3^{ij} (\bx-\by),
\end{equation}
where the delta function is defined in terms of the complete set of transverse vector harmonics as
\begin{equation}
     \dl_3^{ij} (\bx -\by) 
     = \sum_{J \geq \half} \sum_{M,y} Y^{i*}_{J(My)}(\bx)Y^j_{J(My)}(\by).
\end{equation}
The commutation relation for each mode then becomes
\begin{equation}
    [ q_{J_1 (M_1 y_1)}, q^\dag_{J_2 (M_2 y_2)} ] = \dl_{J_1 J_2}\dl_{M_1 M_2}\dl_{y_1 y_2},
\end{equation}
and the Hamiltonian is given by
\begin{eqnarray}
   H^A &=& \int_{S^3} d\Om_3 :\left\{ \half P^i_A P^A_i 
                    - \half A^i \left( \Box_3-2 \right) A_i \right\}:
            \nonumber \\
       &=& \sum_{J \geq \half} \sum_{M,y} (2J+1) q^\dag_{J(My)} q_{J(My)} .
       \label{gauge field Hamiltonian}
\end{eqnarray}

\subsection{Gravitational fields}
\noindent

In order to treat the Weyl action, we decompose the traceless tensor mode as
\begin{equation}
     h_{00} = h,   \qquad  h_{0i} = h_i, \qquad
     h_{ij} = h^{\bf tr}_{ij} + \fr{1}{3} \hgm_{ij} h  , 
\end{equation}
where $h^{\bf tr}_{ij}$ is the spatial component satisfying the traceless condition $h^{{\bf tr} i}_{~~~i}=0$. Then, the gauge transformation (\ref{traceless mode gauge transformation}) is decomposed as
\begin{eqnarray}
    \dl_\kappa h &=& \fr{3}{2} \pd_\eta \kappa_0 + \half \hnb_k \kappa^k, 
        \nonumber \\
    \dl_\kappa h_i &=& \pd_\eta \kappa_i + \hnb_i \kappa_0, 
        \nonumber \\
    \dl_\kappa h^{\bf tr}_{ij} &=& \hnb_i \kappa_j + \hnb_j \kappa_i 
                                   - \fr{2}{3} \hgm_{ij} \hnb_k \kappa^k.
\end{eqnarray}

Using the four gauge degrees of freedom $\kappa^\mu$, we take the transverse gauge defined by the conditions
\begin{equation}
    \hnb^i h_i =\hnb^i h^{\bf tr}_{ij}=0.
       \label{transverse gauge}
\end{equation}
Introducing the transverse vector field $h^\T_i$ and transverse-traceless field $h^{\T\T}_{ij}$, the gauge conditions can be written as
\begin{equation}
   h_i=h_i^\T, \qquad h^{\bf tr}_{ij}=h_{ij}^{\T\T}.
\end{equation}
Then, the combined system of the Wess-Zumino and the Weyl actions (\ref{4DQG action}) on $R\times S^3$ is written in the transverse gauge as
\begin{eqnarray}
    I_{\rm CFT} &=& \int d\eta \int_{S^3} d\Om_3 \biggl\{ 
     -\fr{2b_1}{(4\pi)^2}\phi \left( 
          \pd_\eta^4 -2\Box_3 \pd_\eta^2 +\Box_3^2 + 4\pd_\eta^2 \right) \phi 
                 \nonumber  \\ 
     && \qquad\qquad
        -\fr{1}{2}h^{\T\T}_{ij} \left( \pd_\eta^4 -2\Box_3 \pd_\eta^2 + \Box_3^2 
                     + 8\pd_\eta^2-4\Box_3 +4 \right) h_{\T\T}^{ij} 
                 \nonumber    \\
     && \qquad\qquad
       + h^\T_i \left( \Box_3 +2 \right) 
          \left( -\pd_\eta^2 +\Box_3 -2 \right) h^i_\T  
                \nonumber \\
     && \qquad\qquad
       -\fr{1}{27} h \left( 16 \Box_3 +27 \right) \Box_3 h
            \biggr\} .
          \label{gauge-fixed action}                           
\end{eqnarray}

Since the kinetic term of the $h$ field does not have its time-derivative, this mode is not dynamical. So, using the residual gauge symmetry preserving the transverse-gauge condition, we take the radiation gauge defined by
\begin{equation}
     h=0.
      \label{h=0 gauge}
\end{equation}
Furthermore, since the mode of transverse vector field satisfying the equation $(\Box_3 +2)h^\T_i=0$, which is denoted by the $J=1/2$ vector harmonics, is not dynamical, we remove it as
\begin{equation}
       h^\T_i |_{J=\half}=0.
       \label{J=1/2 gauge}
\end{equation}
We call this choice of the radiation gauge as the radiation$^+$ gauge \cite{hh}. The residual gauge symmetry in the radiation$^+$ gauge is equal to the conformal symmetry, which is discussed in the next section again.

The fourth-order gravitational fields are quantized following the Dirac's procedure \cite{dirac}. Let us first quantize the conformal mode. Introducing the new variable $\chi=\pd_\eta \phi$, we rewrite the action in the second order form
\begin{equation}
     I_\phi =  \int d\eta \int_{S^3} d\Om_3 \left\{ 
           -\fr{b_1}{8\pi^2} \left[
          (\pd_\eta \chi)^2 + 2\chi \Box_3 \chi -4 \chi^2  + ( \Box_3 \phi )^2 
              \right]
            + \upsilon  (\pd_\eta \phi -\chi) \right\},
\end{equation}
where the $\upsilon $ field in the last term is the Lagrange multiplier. The Poisson brackets are then set as 
\begin{eqnarray}
    && \{ \chi(\eta,\bx), P_\chi(\eta, \by) \}_{\rm P} 
       = \{ \phi(\eta,\bx), P_\phi(\eta, \by) \}_{\rm P}
         \nonumber \\
    && = \{ \upsilon(\eta,\bx), P_\upsilon(\eta, \by) \}_{\rm P} 
       = \dl_3 (\bx-\by),
\end{eqnarray}
where the conjugate momenta for $\chi$, $\phi$ and $\upsilon$ are denoted as $P_\chi$, $P_\phi$ and $P_\upsilon$, respectively.

Since $\chi$ is of the second order, it has the ordinary momentum $P_\chi=-(b_1/4\pi^2)\pd_\eta \chi$, while $\phi$ and $\upsilon$ are of the first order and the zeroth order, respectively, and thus they give the constraints
\begin{equation}
    \vphi_1 = P_\phi - \upsilon \simeq 0, \qquad \vphi_2 = P_\upsilon \simeq 0.
\end{equation}
The constraints define the submanifold of the phase space spanned by the six variables of $\chi$, $\phi$, $\upsilon$ and their conjugate momenta, and the weak equalities imply that they are realized on the submanifold.

The Poisson brackets among these constraints are given by 
\begin{equation}
    C_{ab} = \{ \vphi_a, \vphi_b \}_{\rm P} = 
             \left( \begin{array}{cc}
              0 &  -1   \\
              1 &  0 
            \end{array} \right) ,
\end{equation}
where $a,b=1,2$ and the delta function is denoted by $1$. Since $\det C_{ab} \neq 1$, these $\vphi_a$'s are the second class constraints. In order to treat these constraints, we introduce the Dirac bracket defined by
\begin{equation}
   \{ F,G \}_{\rm D} = \{ F,G \}_{\rm P} - \{ F, \vphi_a \}_{\rm P} C^{-1}_{ab} 
                       \{ \vphi_b, G \}_{\rm P}.
\end{equation}
The Dirac bracket has the same properties as the Poisson bracket has.  Since the constraints satisfy the equation $\{ F,\vphi_a \}_{\rm D}=0$ for arbitrary $F$, the Dirac bracket is identified with the Poisson bracket on the submanifold. Taking the Hamiltonian for instance, it implies that the constraints are preserved under the motion on the submanifold.

The Dirac brackets for the four variables on the submanifold are given by
\begin{equation}
   \{ \chi(\eta,\bx), P_\chi(\eta, \by) \}_{\rm D}
     = \{ \phi(\eta,\bx), P_\phi(\eta, \by) \}_{\rm D} 
       = \dl_3 (\bx-\by)
\end{equation}
and the Hamiltonian is written as
\begin{equation}
    H^\phi = \int d\Om_3 \left\{ 
         -\fr{2\pi^2}{b_1}P_\chi^2 + P_\phi \chi 
         + \fr{b_1}{8\pi^2} \left[ 2 \chi \Box_3 \chi -4 \chi^2 + (\Box_3 \phi)^2 \right] 
            \right\}.
        \label{classical Hamiltonian}
\end{equation}
The equations of the motion are then given by 
\begin{eqnarray}
    \pd_\eta \phi &=& \{ \phi, H^\phi \}_{\rm D} = \chi,
            \nonumber \\
    \pd_\eta \chi &=& \{ \chi, H^\phi \}_{\rm D} = - \fr{4\pi^2}{b_1} P_\chi,
            \nonumber \\
    \pd_\eta P_\chi &=& \{ P_\chi, H^\phi \}_{\rm D} 
                  = - P_\phi - \fr{b_1}{2\pi^2} \Box_3 \chi + \fr{b_1}{\pi^2} \chi,
            \nonumber \\
    \pd_\eta P_\phi &=& \{ P_\phi, H^\phi \}_{\rm D} = - \fr{b_1}{4\pi^2} \Box_3^2 \phi.
         \label{H equation of motion}
\end{eqnarray}
The canonical quantization is completed by replacing the Dirac brackets with the commutators,
\begin{equation}
   [ \chi(\eta,\bx), P_\chi(\eta, \by) ]
     = [ \phi(\eta,\bx), P_\phi(\eta, \by) ]
       = i\dl_3 (\bx-\by) . 
       \label{phi commutator}
\end{equation}

The conformal-mode field is expanded in scalar harmonics as $\phi \propto e^{-i\om\eta}Y_{JM}$, and the equation of motion obtained from the action (\ref{gauge-fixed action}), or (\ref{H equation of motion}), is written as
\begin{equation}
     \{ \om^2-(2J)^2 \}\{ \om^2 -(2J+2)^2 \} \phi =0.
\end{equation}
Solving the equation of motion, we write the field as
\begin{eqnarray}
   \phi &=& \fr{\pi}{2\sq{b_1}} \biggl\{ 2({\hat q}+{\hat p}\eta ) Y_{00} 
                \nonumber \\
        && + \sum_{J \geq \half} \sum_M \fr{1}{\sq{J(2J+1)}} 
              \left( a_{JM}e^{-i2J\eta} Y_{JM} 
               + a^\dag_{JM} e^{i2J\eta}Y_{JM}^* \right)
                 \nonumber \\
        && + \sum_{J \geq 0} \sum_M \fr{1}{\sq{(J+1)(2J+1)}} 
           \left( b_{JM}e^{-i(2J+2)\eta} Y_{JM} 
               + b^\dag_{JM} e^{i(2J+2)\eta}Y_{JM}^* \right)  
          \biggr\},
               \nonumber \\
\end{eqnarray}
where $Y_{00}=1/\sq{\V3}=1/\sq{2\pi}$. Calculating the field variables $\chi$, $P_\chi$ and $P_\phi$ using the equations of motion (\ref{H equation of motion}) and setting the equal-time commutation relations (\ref{phi commutator}), we obtain the commutation relation for each mode as
\begin{equation}
    [{\hat q}, {\hat p}]=i, 
    \quad [a_{J_1 M_1},a^\dag_{J_2 M_2}]=\dl_{J_1 J_2}\dl_{M_1 M_2},
    \quad [b_{J_1 M_1},b^\dag_{J_2 M_2}]=-\dl_{J_1 J_2}\dl_{M_1 M_2}.
\end{equation}
Here, $a_{JM}$ and $b_{JM}$ are the positive-metric and the negative-metric operators, respectively.

The Hamiltonian is calculated from expression (\ref{classical Hamiltonian}). Taking the normal ordering we obtain
\begin{equation}
   H^\phi = \half {\hat p}^2 + b_1 + \sum_{J \geq 0} \sum_M 
           \{ 2J a^\dag_{JM} a_{JM} -(2J+2)b^\dag_{JM} b_{JM} \} ,
        \label{conformal mode Hamiltonian}
\end{equation}
up to the constant shift $b_1$. This energy shift is the Casimir effect depending on the coordinate system. Here, it has been determined requiring that the generator of the conformal symmetry on $R \times S^3$, defined in the next section, forms a closed algebra quantum mechanically.

The quantization of the transverse-traceless field $h^{\T\T}_{ij}$ is carried out as in the case of the conformal mode by introducing new variables written in the second order form, while the transverse field $h^\T_i$ is the second order and thus it is quantized in the standard manner.  These fields are expanded in tensor and vector harmonics as $h_{\T\T}^{ij} \propto e^{-i\om\eta}Y^{ij}_{J(Mx)}$ and $h^i_\T \propto e^{-i\om\eta}Y^i_{J(My)}$, respectively.  From the gauge-fixed action (\ref{gauge-fixed action}), the equations of motion are given by 
\begin{eqnarray}
      \{ \om^2-(2J)^2 \}\{ \om^2 -(2J+2)^2 \} h^{ij}_{\T\T} &=& 0,
              \nonumber \\
      (2J-1)(2J+3) \{ \om^2 -(2J+1)^2 \} h^i_\T &=& 0.
\end{eqnarray}
Thus, these fields are expanded as
\begin{eqnarray}
   h_{\T\T}^{ij} &=&
       \fr{1}{4} \sum_{J \geq 1}\sum_{M,x} \fr{1}{\sq{J(2J+1)}} \left\{
            c_{J(Mx)} e^{-i2J\eta}Y^{ij}_{J(Mx)}
            + c^{\dag}_{J(Mx)} e^{i2J\eta}Y^{ij*}_{J(Mx)} \right\}
              \nonumber   \\
   && + \fr{1}{4} \sum_{J \geq 1}\sum_{M,x} \fr{1}{\sq{(J+1)(2J+1)}} \Bigl\{
            d_{J(Mx)} e^{-i(2J+2)\eta}Y^{ij}_{J(Mx)}
                \nonumber   \\
   && \qquad\qquad\qquad\qquad\qquad\qquad\qquad
            + d^{\dag}_{J(Mx)} e^{i(2J+2)\eta}Y^{ij*}_{J(Mx)} \Bigr\},
                \nonumber  \\
    h^i_\T &=&
       \fr{1}{2}\sum_{J \geq 1}\sum_{M,y} \fr{i}{\sq{(2J-1)(2J+1)(2J+3)}}
           \Bigl\{  e_{J(My)} e^{-i(2J+1)\eta}Y^i_{J(My)}
                     \nonumber \\
    && \qquad\qquad\qquad\qquad\qquad\qquad\qquad
          - e^{\dag}_{J(My)} e^{i(2J+1)\eta}Y^{i*}_{J(My)} \Bigr\}.
              \label{mode expansion of traceless mode}
\end{eqnarray}
The commutation relation for each mode is then given by
\begin{eqnarray}
   && \left[c_{J_1 (M_1 x_1)}, c^{\dag}_{J_2 (M_2 x_2)} \right]
      = -\left[d_{J_1 (M_1 x_1)}, d^{\dag}_{J_2 (M_2 x_2)} \right]
      = \dl_{J_1 J_2} \dl_{M_1 M_2}\dl_{x_1 x_2},
             \nonumber   \\
    && \left[e_{J_1 (M_1 y_1)}, e^{\dag}_{J_2 (M_2 y_2)} \right]
      = -\dl_{J_1 J_2}\dl_{M_1 M_2}\dl_{y_1 y_2},
\end{eqnarray}
and the Hamiltonian in the radiation$^+$ gauge is expanded as
\begin{eqnarray}
   H^h &=& \sum_{J \geq 1}\sum_{M,x} \{ 2J c^{\dag}_{J(Mx)}c_{J(Mx)}
                 -(2J+2)d^{\dag}_{J(Mx)}d_{J(Mx)} \}
                \nonumber  \\
    && -\sum_{J \geq 1}\sum_{M,y} (2J+1) e^{\dag}_{J(My)}e_{J(My)} .
      \label{traceless mode Hamiltonian}
\end{eqnarray}
Here, $c_{J(Mx)}$ has the positive-metric, and $d_{J(Mx)}$ and $e_{J(My)}$ have the negative-metric.

\section{Conformal Algebra on $R \times S^3$}
\setcounter{equation}{0}
\noindent

In section 3, we have seen that the diffeomorphism invariance in quantized gravity is described as the conformal symmetry. We here present the generator of the closed conformal algebra on $R \times S^3$ \cite{amm97,hh}, and then examine a long-standing issue on conformal symmetry that the naive conformal transformation does not preserve the gauge-fixing condition in gauge theories \cite{tmp, fp, kluwer}. We here give the answer in the cases of diffeomorphism symmetry in the radiation$^+$ gauge as well as $U(1)$ gauge symmetry in the radiation gauge on $R \times S^3$.

Corresponding to the conformal Killing vectors $\zeta^\mu$ defined by equation (\ref{conformal Killing equation}), there are 15 generators of the conformal algebra given in terms of the stress tensor for the combined system $I_{\rm CFT}$ as
\begin{equation}
   Q_{\zeta}=\int_{S^3} d\Om_3 \zeta^{\mu} :\hat{T}_{\mu 0}:
           \label{conformal charge}
\end{equation}
where the stress tensor is defined by the variation with respect to the background metric as 
\begin{equation}
   \hat{T}^{\mu\nu}=\fr{2}{\sq{-\hg}}\fr{\dl I_{\rm CFT}}{\dl \hg_{\mu\nu}} 
\end{equation}
and $\hat{T}_{\mu\nu}=\hg_{\mu\lam}\hg_{\nu\s}\hat{T}^{\lam\s}$ satisfying the traceless condition $\hat{T}^\lam_{~\lam}=0$.

The conformal Killing equation (\ref{conformal Killing equation}) is written in components as
\begin{eqnarray}
     3 \pd_\eta \zeta_0 + \psi &=& 0, 
       \nonumber \\
     \pd_\eta \zeta_i + \hnb_i \zeta_0 &=& 0,
       \nonumber \\
     \hnb_i \zeta_j + \hnb_j \zeta_i - \fr{2}{3} \hgm_{ij} \psi &=& 0
       \label{component of conformal Killing equation}
\end{eqnarray}
where $\psi=\hnb_i \zeta^i$. Using these equations and the conservation equation of the stress tensor, we can show that the generator is conserved: 
\begin{equation}
     \fr{d Q_\zeta}{d\eta} = -\fr{1}{3}\int_{S^3} d\Om_3 \psi \hat{T}^\lam_{~\lam}=0.
\end{equation}

Solving the conformal Killing equations with respect to $\psi$, we obtain the equations
\begin{equation}
    (\Box_3+3)\psi=0, \qquad (\pd_\eta^2 +1 )\psi =0.
\end{equation}
The left equation is derived by acting the operator $\hnb^j \hnb^i$ to the last equation in (\ref{component of conformal Killing equation}). Combining the left equation and other conformal Killing equations, we obtain the right equation. Thus, $\psi$ is given by
\begin{equation}
     \psi = 0 \quad \hbox{or} \quad \psi \propto e^{\pm i\eta}Y_{\half M}.
\end{equation}

First, we consider the case of $\psi=0$. This solution satisfies $\pd_\eta \zeta_0=\Box_3 \zeta_0=0$ and the Killing equation $\hnb_i \zeta_j + \hnb_j \zeta_i =0$. One of them satisfying these conditions is the constant Killing vector generating the time evolution,
\begin{equation}
    \zeta^\mu_\T=(1,0,0,0). 
\end{equation}
The other is the Killing vector generating the isometry group $SU(2)\times SU(2)$ on $S^3$, which satisfies the Killing equation and the conditions $\zeta_0=\pd_\eta \zeta_i=0$. The Killing vector on $S^3$ can be written in terms of the $J=1/2$ scalar harmonics as $\zeta_{\rm R}^\mu=(0,\zeta_{\rm R}^i)$ with
\begin{equation}
    ( \zeta^i_{\rm R} )_{MN} 
      = i \fr{\V3}{4} \left\{ Y^*_{\half M} \hnb^i Y_{\half N} 
              - Y_{\half N} \hnb^i Y^*_{\half M}  \right\} .
\end{equation}
This vector can be rewritten in terms of the transverse vector harmonics with $J=1/2$ (see equation (\ref{Killing vector in terms of J=1/2 vector harmonics})).

Substituting these Killing vectors into the definition (\ref{conformal charge}), we obtain the Hamiltonian
\begin{equation}
         H=\int_{S^3} d\Om_3 :\hat{T}_{00}:
\end{equation}
and 6 generators of the rotation group $SU(2) \times SU(2)$, 
\begin{equation}
     R_{MN} =\int_{S^3} d\Om_3 (\zeta^i_{\rm R})_{MN} :\hat{T}_{i0}: 
          \label{rotation generator}
\end{equation}
with the properties
\begin{equation}
     R_{MN}=-\eps_M \eps_N R_{-N-M}, \qquad  R^\dag_{MN}=R_{NM} . 
\end{equation}

The solution of the conformal Killing equation with $\psi \neq 0$ is given by 
\begin{equation} 
   (\zeta_\S^0)_M = \half \sq{\V3} e^{i\eta}Y^*_{\half M} , \quad 
   (\zeta_\S^i)_M = -\fr{i}{2} \sq{\V3} e^{i\eta} \hnb^i Y^*_{\half M} 
          \label{zeta_S vector}
\end{equation}
and its complex conjugate. Substituting this vector into the definition and rewriting it using the conservation equation of the stress tensor, we obtain the following expression:
\begin{equation}
    Q_M = \sq{\V3} P^{(+)} \int_{S^3} d\Om_3 
            Y^*_{\half M} :\hat{T}_{00}: , 
            \label{Q_M charge}
\end{equation}
where $P^{(+)}=e^{i\eta}(1+i\pd_\eta)/2$. The integration over $S^3$ selects out the terms with the phase factor $e^{\pm i\eta}$, and the projection operator $P^{(+)}$ chooses the $e^{-i\eta}$ part and make the generator time-independent. The generator $Q_M$ and its hermite conjugates $Q_M^\dag$ are the $4+4$ generators of the special conformal transformations (precisely proper combinations of the translation and the special conformal transformation).

We here reconsider the radiation$^+$ gauge. The space of the residual diffeomorphism symmetry preserving the radiation gauge conditions, (\ref{transverse gauge}) and (\ref{h=0 gauge}), are defined by the equations, $\dl_\kappa h=(3\pd_\eta \kappa_0+\tilde{\psi})/2=0$, $\dl_\kappa (\hnb_i h^i)=\pd_\eta \tilde{\psi} + \Box_3 \kappa_0 =0$ and $\dl_\kappa (\hnb^i h^{\bf tr}_{ij})= (\Box_3 +2)\kappa_j + \hnb_j \tilde{\psi}/3 =0$, where $\tilde{\psi}=\hnb_i \kappa^i$. The space of the residual symmetry in the radiation gauge is bigger than the space generated by the 15 conformal Killing vectors. The second equation shows that there is the Killing vector on $S^3$ satisfying equation $\pd_\eta \kappa^i \neq 0$ represented by $\kappa^\mu=(0,fY^i_{1/2(My)})$ for arbitrary function of time, $f$. So, using this residual gauge degree of freedom we can remove the $J=1/2$ mode in the transverse vector field $h^i_\T$ (\ref{J=1/2 gauge}). This is the radiation$^+$ gauge, and then the residual diffeomorphism symmetry becomes the conformal symmetry generated by the 15 conformal Killing vectors.

The 15 generators of the conformal symmetry form the closed algebra of $SO(4,2)$:
\begin{eqnarray}
     \left[ Q_M, Q^\dag_N \right] &=& 2\dl_{MN} H + 2R_{MN},
           \nonumber \\
    \left[ H, Q_M \right] &=& -Q_M,
           \nonumber \\
    \left[ H, R_{MN} \right] &=& \left[ Q_M, Q_N \right] = 0,
           \nonumber  \\
    \left[ Q_M, R_{M_1 M_2} \right] &=& \dl_{M M_2}Q_{M_1}
                 -\eps_{M_1}\eps_{M_2}\dl_{M -M_1}Q_{-M_2} ,
            \nonumber  \\
    \left[ R_{M_1 M_2}, R_{M_3 M_4} \right]
        &=& \dl_{M_1 M_4} R_{M_3 M_2} 
            -\eps_{M_1}\eps_{M_2} \dl_{-M_2 M_4} R_{M_3 -M_1}
            \nonumber \\
    && - \dl_{M_2 M_3} R_{M_1 M_4} 
       +\eps_{M_1}\eps_{M_2} \dl_{-M_1 M_3} R_{-M_2 M_4} .
            \label{conformal algebra}
\end{eqnarray}
The Hamiltonian on the cylindrical background $R \times S^3$ is the dilatation operator counting the conformal weight of the state.\footnote{ 
It can be seen considering conformal map $y \to r = e^y$ from the Euclidean $R\times S^3$ with the metric $dy^2 + d\Om^2_3$ to $R^4$ with the metric $dr^2 +r^2 d\Om^2_3$. The dilatation $r \to e^a r$ is just the time translation $y \to y +a$ on the cylindrical manifold. The procedure for defining a quantum theory on $R^4$ is known as radial quantization \cite{fhj}. A quantum theory on the metric of $R\times S^3$ (\ref{metric}) is yielded by the analytical continuation, $y=i\eta$. Therefore, each mode with the time dependence $e^{iE\eta}$ has the conformal weight $E$.
} 
The rotation generator $R_{MN}$ has the vanishing conformal weight and has a diagonal form for each mode labeled by $J$. On the other hand, the generator of the special conformal transformation $Q_M$ has the weight $-1$ (its conjugate has the weight $1$), and it is the ${\bf 4}$-vector on $SU(2)\times SU(2)$. Thus, this generator is given by a ${\bf 4}$-vector constructed as a proper combination of creation and annihilation modes with different weights by $1$.

The rotation algebra can be rewritten in the familiar form of the $SU(2)\times SU(2)$ algebra. Parametrizing the ${\bf 4}$-vector $\{ (\half, \half), (\half, -\half), (-\half, \half),  (-\half, -\half) \}$ by $\{ 1,2,3,4 \}$, and setting $A_+=R_{31}$, $A_-=R_{31}^\dag$, $A_3=(R_{11}+R_{22})/2$, $B_+=R_{21}$, $B_-=R_{21}^\dag$ and $B_3=(R_{11}-R_{22})/2$, the last algebra in (\ref{conformal algebra}) is written in the form
\begin{eqnarray}
  &&  [A_+, A_-]=2A_3, \qquad [A_3, A_\pm ]=\pm A_\pm,
         \nonumber \\
  &&  [B_+, B_-]=2B_3, \qquad [B_3, B_\pm ]=\pm B_\pm,
\end{eqnarray}
where $A_{\pm,3}$ and $B_{\pm,3}$ commute. The generators $A_{\pm,3} (B_{\pm,3})$ act on the left (right) index of $M=(m,m^\pp)$.

The four-dimensional quantum gravity is now decomposed into four sectors: the scalar field, the gauge field, the conformal mode and the traceless tensor mode. The full generator is given by combining all sectors as
\begin{equation}
    Q_\zeta = Q_\zeta^X + Q_\zeta^A + Q_\zeta^\phi + Q_\zeta^h .
\end{equation}

\subsection{Scalar fields}
\noindent

Let us first construct the generator of the conformal algebra for the scalar field. The stress tensor is given by 
\begin{equation}
    \hat{T}^X_{\mu\nu} = \fr{2}{3}\hnb_\mu X \hnb_\nu X -\fr{1}{3} X \hnb_\mu \hnb_\nu X 
                       - \fr{1}{6}\hg_{\mu\nu} \left\{ \hnb_\lam X \hnb^\lam X 
                            + \fr{1}{6}\hR X^2 \right\} + \fr{1}{6} \hR_{\mu\nu} X^2 .
\end{equation}
The trace of the stress tensor vanishes in proportion to the equation of motion as
\begin{equation}
     \hat{T}^{X \lam}_{~~~\lam} = \fr{1}{3}X \left( -\hnb^2 + \fr{1}{6}\hR \right)X =0,
\end{equation}
and thus the generator is conserved.

The Hamiltonian is given by $H^X$ (\ref{scalar field Hamiltonian}) and the generator of the special conformal transformation is calculated by using expression (\ref{Q_M charge}) as \cite{amm97}
\begin{eqnarray}
    Q^X_M &=& \sum_{J \geq 0}\sum_{M_1}\sum_{M_2} \C^{\half M}_{JM_1, J+\half M_2}  
           \sq{(2J+1)(2J+2)}  \eps_{M_1}\vphi^{\dag}_{J-M_1}\vphi_{J+\half M_2} .
        \nonumber \\
\end{eqnarray}
The $\C$ function is the $SU(2)\times SU(2)$ Clebsch-Gordan coefficient defined by the integral of three products of scalar harmonics over $S^3$,
\begin{eqnarray}
     \C^{JM}_{J_1M_1,J_2M_2}
      &=& \sq{\V3} \int_{S^3} d\Om_3 Y^*_{JM}Y_{J_1M_1}Y_{J_2M_2}
             \nonumber \\
      &=& \sq{\fr{(2J_1+1)(2J_2+1)}{2J+1}} C^{Jm}_{J_1m_1,J_2m_2}
                           C^{J\prm}_{J_1\prm_1,J_2\prm_2} ,
        \label{C function}
\end{eqnarray} 
where $C^{Jm}_{J_1 m_1, J_2 m_2}$ is the standard $SU(2)$ Clebsch-Gordan coefficient \cite{vmk}.

In order to obtain the rotation generator, we write the Killing vector using the $J=1/2$ vector harmonics as 
\begin{equation}
    \left( \xi_{\rm R}^i \right)_{MN} 
        = i \half \sq{\V3} \sum_{V,y}
               \G^{\half M}_{\half (Vy); \half N} Y^{i*}_{\half (Vy)} . 
          \label{Killing vector in terms of J=1/2 vector harmonics}
\end{equation}
The $\G$ function is the $SU(2)\times SU(2)$ Clebsch-Gordan coefficient defined by the integral \cite{hh}
\begin{equation}
     \G^{JM}_{J_1(M_1 y_1);J_2 M_2}
     = \sq{\V3} \int_{S^3} d\Om_3 Y^*_{JM} Y^i_{J_1 (M_1y_1)} \hnb_i Y_{J_2 M_2} .
\end{equation}
Using this function, the rotation generator for the scalar field can be written as
\begin{equation}
    R^X_{MN} = -\half \sum_{J \geq 0} \sum_{S_1}\sum_{S_2} \sum_{V,y}(- \eps_V) 
              \G^{\half M}_{\half (-Vy); \half N} \G^{J S_1}_{\half (Vy); J S_2}
              \vphi^\dag_{J S_1} \vphi_{J S_2} .
\end{equation}
Substituting the concrete forms of the $\G$ functions,
\begin{eqnarray}
      \G^{\half M}_{J (Vy);JN} &=& -\hbox{$\sq{2J(2J+2)}$}
          C^{\half m}_{J+y v, Jn}C^{\half m^\pp}_{J-y v^\pp,Jn^\pp},
             \nonumber  \\
   \G^{JM}_{\half (Vy);JN} &=& -\hbox{$\sq{2J(2J+2)}$}
          C^{Jm}_{\half+y v, Jn}C^{Jm^\pp}_{\half-y v^\pp,Jn^\pp},
\end{eqnarray}
we obtain the following form:
\begin{eqnarray}
   R^X_{11} &=& \sum_{J > 0}\sum_M (m+\prm)  
             \vphi^\dag_{JM}\vphi_{JM} , 
           \nonumber   \\
   R^X_{22} &=& \sum_{J > 0}\sum_M (m-\prm)  
             \vphi^\dag_{JM}\vphi_{JM} ,   
           \nonumber   \\
   R^X_{21} &=& \sum_{J > 0}\sum_M \sq{(J-\prm+1)(J+\prm)} 
             \vphi^\dag_{JM}\vphi_{J \overline{M}} , 
           \nonumber  \\ 
   R^X_{31} &=& \sum_{J > 0}\sum_M \sq{(J-m+1)(J+m)}  
             \vphi^\dag_{JM}\vphi_{J \underline{M}} ,
\end{eqnarray}
where the indices with over and under lines are defined by $\overline{M}=(m,m^\pp-1)$ and $\underline{M}=(m-1,m^\pp)$, respectively.

The conformal transformation of the scalar field (\ref{scalar field conformal transformation}) is then expressed in terms of the commutator as
\begin{equation}
     \dl_\zeta X = i[Q^X_\zeta,X] .
\end{equation}
It is known that the time translation $\dl_{\zeta_\T}X=\pd_\eta X$ is expressed by the commutator $i[H^X,X]$. For the case of the special conformal transformation, we can directly see that the transformation (\ref{scalar field conformal transformation}) with $\zeta_\S^\mu$ (\ref{zeta_S vector}) is expressed by the commutator $i[Q^X_M,X]$ using the product expansions
\begin{eqnarray}
    && Y_{\half M}^* Y_{JN} = \fr{1}{\sq{\V3}}
             \left\{ \sum_S \C^{\half M}_{JN,J+\half S}Y^*_{J+\half S}
                     + \sum_S \C^{\half M}_{JN,J-\half S}Y^*_{J-\half S} \right\},
                    \nonumber \\
   && \hnb^i Y_{\half M}^* \hnb_i Y_{JN} = \fr{1}{\sq{\V3}}
             \biggl\{ -2J\sum_S \C^{\half M}_{JN,J+\half S}Y^*_{J+\half S}
             \nonumber \\
   && \qquad\qquad\qquad\qquad\qquad
             +(2J+2) \sum_S \C^{\half M}_{JN,J-\half S}Y^*_{J-\half S} \biggr\} . 
      \label{product expansion for scalar conformal transformation}
\end{eqnarray}

\subsection{Gauge fields}
\noindent

The stress tensor for the gauge field is given by
\begin{equation}
   \hat{T}^A_{\mu\nu} = F_{\mu\lam}F^{~\lam}_\nu - \fr{1}{4} \hg_{\mu\nu} F_{\lam\s}F^{\lam\s}.
\end{equation}
where the space-time index of the gauge field is raised using the background metric as $F^\mu_{~\nu}=\hg^{\mu\lam}F_{\lam\nu}$. The trace of the stress tensor trivially vanishes, and thus the generator of the conformal algebra is conserved.

We here give the generators in the radiation gauge defined by $A_0=\hnb^i A_i=0$, (\ref{Coulomb gauge}) and (\ref{A_0=0 gauge}). The Hamiltonian $H^A$ has already calculated in (\ref{gauge field Hamiltonian}). The generator of the special conformal transformation is given by \cite{hh}
\begin{eqnarray}
    Q^A_M &=& \sum_{J \geq \half}\sum_{M_1,y_1}\sum_{M_2,y_2} 
        \D^{\half M}_{J(M_1 y_1), J+\half (M_2 y_2)}  
           \sq{(2J+1)(2J+2)}  
        \nonumber \\
          && \qquad\qquad\qquad\qquad \times 
           (-\eps_{M_1}) q^{\dag}_{J(-M_1 y_1)}q_{J+\half (M_2 y_2)} ,
\end{eqnarray}
where the $\D$ function is the $SU(2) \times SU(2)$ Clebsch-Gordan coefficient defined by the integral of the product of one scalar and two vector harmonics over $S^3$ \cite{hh}. We here write the expression for the special case of the $\D$ function that appears in the generator, 
\begin{eqnarray}
       \D^{\half M}_{J(M_1 y_1), J+\half (M_2 y_2)}
       &=& \sq{\V3}
         \int_{S^3} d\Om_3 Y^*_{\half M}Y^i_{J(M_1y_1)}Y_{i J+\half(M_2y_2)}
              \nonumber \\
       &=&  \sq{J(2J+3)}  C^{\half m}_{J+y_1 m_1, J+\half+y_2 m_2} 
              C^{\half \prm}_{J-y_1 \prm_1, J+\half-y_2 \prm_2} .
                \nonumber \\ 
       &&  \label{D function}
\end{eqnarray}
The rotation generator is not depicted here because we do not use the explicit form of it below.

Now, we discuss the issue on the conformal invariance in gauge theories. The conformal transformation for the spatial component of the gauge field is written in the radiation gauge as
\begin{equation}
     \dl_\zeta A_i = \zeta^0 \pd_\eta A_i + \zeta^j \hnb_j A_i 
              + \fr{1}{3} \psi A_i 
              + \half \left( \hnb_i \zeta^j - \hnb^j \zeta_i \right) A_j .
       \label{gauge fixed conformal transformation}
\end{equation}
The gauge-fixed action is invariant under this naive conformal transformation. However, this transformation does not preserve the transverse gauge condition (\ref{Coulomb gauge}) for the special conformal transformation, $\dl_{\zeta_\S}$. The time component of the gauge field also transforms as 
\begin{equation}
    \dl_\zeta A_0 = \hnb^i (\zeta^0 A_i) .
      \label{gauge fixed conformal transformation for A_0}
\end{equation}
Thus, the radiation gauge is not preserved under the special conformal transformation.

Since the time translation and the rotation preserve the gauge conditions, we focus on the special conformal transformation below. Using the product expansions  
\begin{eqnarray}
    && Y_{\half M}^* Y^i_{J(Ny)}
          \nonumber \\
    && = \fr{1}{\sq{\V3}}
         \biggl\{ \sum_{V,y^\pp} \D^{\half M}_{J(Ny),J+\half (Vy^\pp)}Y^{i*}_{J+\half (Vy^\pp)}
               + \sum_{V,y^\pp} \D^{\half M}_{J(Ny),J-\half (Vy^\pp)}Y^{i*}_{J-\half (Vy^\pp)}
             \nonumber \\
     && \qquad\qquad
               + \fr{1}{2J(2J+2)} \sum_S \G^{\half M}_{J(Ny);JS} \hnb^i Y^*_{JS} 
               \biggr\},
                   \nonumber \\
   && \hnb^j Y_{\half M}^* \hnb_j Y^i_{J(Ny)} 
          \nonumber \\
   && = \fr{1}{\sq{\V3}}
       \biggl\{ -2J\sum_{V,y^\pp} \D^{\half M}_{J(Ny),J+\half (Vy^\pp)}Y^{i*}_{J+\half (Vy^\pp)}
             \nonumber \\
   && \qquad\qquad
          +(2J+2) \sum_{V,y^\pp} \D^{\half M}_{J(Ny),J-\half (Vy^\pp)}Y^{i*}_{J-\half (Vy^\pp)} 
             \nonumber \\
    && \qquad\qquad
          + \fr{2}{2J(2J+2)} \sum_S \G^{\half M}_{J(Ny);JS} \hnb^i Y^*_{JS}
          \biggr\} ,
\end{eqnarray}
we can show that the conformal transformation in the radiation gauge (\ref{gauge fixed conformal transformation}) with the conformal Killing vector $\zeta^\mu=\zeta_\S^\mu$ is written as
\begin{equation}
     \dl_{\zeta_\S} A_i = i[Q^A_M, A_i]+ \hnb_i \lam_\S .
\end{equation}
The second term in the right-hand side breaks the transverse gauge conditions, which is given by the mode-dependent scalar function
\begin{eqnarray}
   && (\lam_\S)_M = \fr{i}{2} \sum_{J \geq \half} \fr{1}{\sq{2(2J+1)}}\sum_{N,y} \sum_S 
             \biggl\{ -\fr{1}{2J} q_{J(Ny)}e^{-i2J\eta}\G^{\half M}_{J(Ny);JS}
               \nonumber \\
   && \qquad\qquad\qquad              
             +\fr{1}{2J+2}q^\dag_{J(Ny)} e^{i(2J+2)\eta}(-\eps_N)\G^{\half M}_{J(-Ny);JS}
               \biggr\} Y^*_{JS} .
\end{eqnarray}

The breaking term has the form of the $U(1)$ gauge transformation. So, we consider the gauge transformation with the parameter $\lam_\S$, 
\begin{equation}
   \dl_{\lam_\S}A_\mu=\hnb_\mu \lam_\S ,
      \label{gauge transformation with lambda_S}
\end{equation}
and rewrite the conformal transformation in the form
\begin{equation}
    \dl_{\zeta_\S}A_i -\dl_{\lam_\S} A_i = i[Q^A_M,A_i] .
\end{equation}
We can then show that the transformation of the time-component field satisfies the equation
\begin{equation}
     \dl_{\zeta_\S}A_0 - \dl_{\lam_\S}A_0 = \hnb^i(\zeta^0_\S A_i)-\pd_\eta \lam_\S =0.
\end{equation}

Thus, the transformation yielded by the generator of the conformal algebra is expressed in terms of the combined transformation
\begin{equation}
    \dl^\T_\zeta =\dl_\zeta - \dl_{\lam_\zeta},
\end{equation}
where the mode-dependent gauge parameter $\lam_\zeta$ is given by $\lam_\S$ for the generator $Q^A_M$ and vanishes for the Hamiltonian and the rotation generator, such that
\begin{eqnarray}
      \dl^\T_\zeta A_i &=& i[Q^A_\zeta A_i],
           \nonumber  \\
      \dl^\T_\zeta A_0 &=& 0 .
\end{eqnarray}
This transformation just forms the closed algebra with preserving the radiation gauge.

\subsection{The conformal mode}
\noindent

Taking the variation of the Wess-Zumino action with respect to the background metric, we obtain the stress tensor for the conformal mode, 
\begin{eqnarray}
    && \hat{T}^\phi_{\mu\nu} = -\fr{b_1}{8\pi^2} \biggl\{ 
      -4 \hnb^2 \phi \hnb_\mu \hnb_\nu \phi + 2 \hnb_\mu \hnb^2 \phi \hnb_\nu \phi
      + 2 \hnb_\nu \hnb^2 \phi \hnb_\mu \phi 
         \nonumber \\
     && \quad
      + \fr{8}{3} \hnb_\mu \hnb_\lam \phi \hnb_\nu \hnb^\lam \phi 
      -\fr{4}{3} \hnb_\mu \hnb_\nu \hnb_\lam \phi \hnb^\lam \phi
      + 4 \hR_{\mu\lam\nu\s}\hnb^\lam \phi \hnb^\s \phi 
        \nonumber \\
     && \quad
      + 4 \hR_{\mu\lam} \hnb^\lam \phi \hnb_\nu \phi 
      + 4 \hR_{\nu\lam} \hnb^\lam \phi \hnb_\mu \phi
      - \fr{4}{3} \hR_{\mu\nu}\hnb_\lam \phi \hnb^\lam \phi 
      - \fr{4}{3} \hR \hnb_\mu \phi \hnb_\nu \phi 
         \nonumber \\
     && \quad
      - \fr{2}{3} \hnb_\mu \hnb_\nu \hnb^2 \phi 
      - 4 \hR_{\mu\lam\nu\s}\hnb^\lam \hnb^\s \phi + \fr{14}{3} \hR_{\mu\nu} \hnb^2 \phi
      + 2 \hR \hnb_\mu \hnb_\nu \phi
         \nonumber \\
     && \quad
      - 4 \hR_{\mu\lam}\hnb^\lam \hnb_\nu \phi  - 4 \hR_{\nu\lam}\hnb^\lam \hnb_\mu \phi
      - \fr{1}{3} \hnb_\mu \hR \hnb_\nu \phi   - \fr{1}{3} \hnb_\nu \hR \hnb_\mu \phi
         \nonumber \\
     && \quad
      + \hg_{\mu\nu} \biggl[ 
         \hnb^2 \phi \hnb^2 \phi   - \fr{2}{3} \hnb^\lam \hnb^2 \phi \hnb_\lam \phi 
         - \fr{2}{3} \hnb^\lam \hnb^\s \phi \hnb_\lam \hnb_\s \phi 
         - \fr{8}{3} \hR_{\lam\s} \hnb^\lam \phi \hnb^\s \phi 
           \nonumber \\
     && \quad
         + \fr{2}{3} \hR \hnb^\lam \phi \hnb_\lam \phi
         + \fr{2}{3} \hnb^4 \phi 
         + 4 \hR_{\lam\s} \hnb^\lam \hnb^\s \phi   - 2 \hR \hnb^2 \phi 
         + \fr{1}{3} \hnb^\lam \hR  \hnb_\lam \phi
         \biggr]  \biggr\}.
            \nonumber \\
\end{eqnarray}
The trace of the stress tensor vanishes in propotion to the equation of motion as
\begin{equation}
    \hat{T}^{\phi \lam}_{~~~\lam} 
    = -\fr{b_1}{4\pi^2} \hDelta_4 \phi  =0,
\end{equation}
where $\hE_4=0$ on $R \times S^3$, and thus the generator of conformal algebra is conserved.

The Hamiltonian $H^\phi$ has been computed in (\ref{conformal mode Hamiltonian}). The generator of the special conformal transformation has the form \cite{amm97}
\begin{eqnarray}
    Q^\phi_M &=& \left( \hbox{$\sq{2b_1}$}-i\hat{p} \right) a_{\half M}
             \nonumber \\
       &&  +\sum_{J \geq 0}\sum_{M_1}\sum_{M_2} \C^{\half M}_{JM_1, J+\half M_2}
              \Bigl\{ \a(J) \eps_{M_1} a^\dag_{J-M_1} a_{J+\half M_2}
                  \nonumber  \\
      && \qquad
             +\b(J) \eps_{M_1} b^\dag_{J-M_1} b_{J+\half M_2}
             +\gm(J) \eps_{M_2} a^\dag_{J+\half -M_2} b_{J M_1} \Bigr\} ,
\end{eqnarray}
where the $\C$ function is defined by equation (\ref{C function}) and the coefficients are given by
\begin{equation}
    \a(J) = \sq{2J(2J+2)},
          \quad
    \b(J) = -\sq{(2J+1)(2J+3)},
            \quad
    \gm(J)= 1  .
        \label{alpha beta gamma}
\end{equation}
The significant property of the generator $Q^\phi_M$ is that this generator mixes the positive-metric mode and the negative-metric mode. The rotation generator is not depicted here because we do not use the explicit form of it below.

We here give a crossing relation among the $SU(2)\times SU(2)$ Clebsch-Gordan coefficients which is useful to check that these generators just form the closed algebra and to obtain physical states in the next section. Consider the integral of four products of scalar harmonics over $S^3$,  
\begin{equation}
   \int_{S^3} d\Om_3 Y^*_{J_1M_1}Y_{J_2M_2}Y^*_{J_3M_3}Y_{J_4M_4} .
         \label{4 scalar product} 
\end{equation}
Applying the product expansion
\begin{equation}
    Y_{J_1 M_1} Y_{J_2 M_2} = \fr{1}{\sq{\V3}} \sum_{J \geq 0} \sum_M
                 \C^{JM}_{J_1 M_1, J_2 M_2} Y_{JM}
\end{equation}
to four products of scalar harmonics, we obtain the crossing relation \cite{hh}
\begin{equation}
    \sum_{J \geq 0} \sum_M \eps_M \C^{J_1M_1}_{J_2M_2, J-M} 
                               \C^{J_3M_3}_{JM,J_4M_4} 
    = \sum_{J \geq 0} \sum_M \eps_M \C^{J_1M_1}_{J_4M_4, J-M} 
                               \C^{J_3M_3}_{JM,J_2M_2}  .                         
           \label{crossing relation}
\end{equation}
Using the crossing property with $J_1=J_3=1/2$ we can reduce the calculation of the commutator $[Q_M^\phi,Q_N^{\phi \dag}]$ in the conformal algebra.

The conformal transformation of the conformal mode (\ref{conformal mode conformal transformation}) is written in terms of the commutator as
\begin{equation}
     \dl_\zeta \phi = i[Q^\phi_\zeta, \phi] .
\end{equation} 
It can be shown by using the product expansion (\ref{product expansion for scalar conformal transformation}) as in the case of the scalar field.

\subsection{The traceless tensor mode}
\noindent

We here study the generator for the traceless tensor mode and the property of the conformal transformation in the radiation$^+$ gauge.

The Hamiltonian $H^h$ has been derived from the Weyl action in (\ref{traceless mode Hamiltonian}). The generator of the special conformal transformation is given by \cite{hh}
\begin{eqnarray}
    Q^h_M &=& \sum_{J \geq 1}\sum_{M_1,x_1}\sum_{M_2,x_2}
            \E^{\half M}_{J(M_1 x_1), J+\half (M_2 x_2)}
          \Bigl\{ \a(J) \eps_{M_1} c^\dag_{J(-M_1 x_1)} c_{J+\half (M_2 x_2)}
                        \nonumber  \\
     && \qquad
             +\b(J) \eps_{M_1} d^\dag_{J(-M_1 x_1)} d_{J+\half (M_2 x_2)}
             +\gm(J) \eps_{M_2} c^\dag_{J+\half (-M_2 x_2)} d_{J (M_1 x_1)} \Bigr\}
                        \nonumber \\
     && + \sum_{J \geq 1}\sum_{M_1,x_1}\sum_{M_2,y_2}
            \H^{\half M}_{J(M_1 x_1); J (M_2 y_2)}
             \Bigl\{ A(J) \eps_{M_1} c^\dag_{J(-M_1 x_1)} e_{J (M_2 y_2)}
                   \nonumber  \\
        &&\qquad\qquad\qquad\qquad\qquad\qquad\qquad
                - B(J) (-\eps_{M_2}) e^\dag_{J(-M_2 y_2)} d_{J (M_1 x_1)} \Bigr\}
                     \nonumber  \\
        && - \sum_{J \geq 1}\sum_{M_1,y_1}\sum_{M_2, y_2}
           \D^{\half M}_{J(M_1 y_1), J+\half (M_2 y_2)}
             C(J) (-\eps_{M_1}) e^\dag_{J(-M_1 y_1)} e_{J+\half (M_2 y_2)} ,
                \nonumber \\
\end{eqnarray}
where the coefficients, $\a(J)$, $\b(J)$ and $\gm(J)$ are equal to those in the generator for the conformal mode (\ref{alpha beta gamma}). Other coefficients are given by
\begin{eqnarray}
    A(J) &=& \sq{\fr{4J}{(2J-1)(2J+3)}},
              \nonumber   \\
    B(J) &=& \sq{\fr{2(2J+2)}{(2J-1)(2J+3)}},
               \nonumber   \\
    C(J) &=& \sq{\fr{(2J-1)(2J+1)(2J+2)(2J+4)}{2J(2J+3)}} .
\end{eqnarray}
The $\E$ and $\H$ functions are the $SU(2)\times SU(2)$ Clebsch-Gordan coefficients defined by \cite{hh}
\begin{eqnarray}
   \E^{\half M}_{J(M_1 x_1), J+\half (M_2 x_2)}
       &=& \sq{\V3}
         \int_{S^3} d\Om_3 Y^*_{\half M}Y^{ij}_{J(M_1x_1)}Y_{ij J+\half(M_2x_2)}
              \nonumber \\
       &=& \sq{(2J-1)(J+2)} C^{\half m}_{J+x_1 m_1, J+\half+x_2 m_2} 
              C^{\half \prm}_{J-x_1 \prm_1, J+\half-x_2 \prm_2} ,
                \nonumber \\ \nonumber \\
   \H^{\half M}_{J(M_1 x_1); J(M_2 y_2)}
       &=& \sq{\V3}
         \int_{S^3} d\Om_3 Y^*_{\half M}Y^{ij}_{J(M_1x_1)}\hnb_iY_{j J(M_2y_2)}
              \nonumber \\
       &=& - \sq{(2J-1)(2J+3)} 
              C^{\half m}_{J+x_1 m_1, J+y_2 m_2} 
              C^{\half \prm}_{J-x_1 \prm_1, J-y_2 \prm_2} ,
                \nonumber \\
\end{eqnarray}
and the $\D$ function is given by (\ref{D function}).

The stress tensor for the Weyl action is quite complicated, and so we have derived the generator of the special conformal transformation indirectly: assume a generic form with arbitrary six coefficients, $\a$, $\b$, $\gm$, $A$, $B$ and $C$, and then determine them imposing the condition that the generator forms the closed algebra of conformal symmetry. The conventions of the mode expansion (\ref{mode expansion of traceless mode}) and the coefficients in the generator are fixed so as to match the conformal transformation discussed below.

We here emphasize that the manner to derive this algebra is generic, and thus the existence of the cross terms of the positive-metric and the negative-metric modes means that the higher-derivative gravitatioal action including the negative-metric modes is required in order for quantum diffeomorphism symmetry to form the closed algebra.

The conformal transformations (\ref{traceless mode conformal transformation}) in the radiation$^+$ gauge are written in components as
\begin{eqnarray}
     \dl_\zeta h^{\T\T}_{ij} &=& 
        \zeta^0 \pd_\eta h^{\T\T}_{ij} + \zeta^k \hnb_k h^{\T\T}_{ij}
        + \half \left( \hnb_i \zeta^k - \hnb^k \zeta_i \right) h^{\T\T}_{jk}
             \nonumber \\
       && + \half \left( \hnb_j \zeta^k - \hnb^k \zeta_j \right) h^{\T\T}_{ik}
          + h^\T_i \hnb_j \zeta^0 + h^\T_j \hnb_i \zeta^0 
           -\fr{2}{3} \gm_{ij}  \hnb^k \left( \zeta^0 h^\T_k \right), 
             \nonumber \\
     \dl_\zeta h^\T_i &=& 
        \zeta^0 \pd_\eta h^\T_i + \zeta^k \hnb_k h^\T_i
        + \half \left( \hnb_i \zeta^k - \hnb^k \zeta_i \right) h^\T_k
        +  \hnb^k \left( \zeta^0 h^{\T\T}_{ik} \right),
            \nonumber \\
     \dl_\zeta h &=& 2 \hnb^k \left( \zeta^0 h^\T_k \right) .          
\end{eqnarray}
These transformations do not preserve the radiation$^+$ gauge for the case of special conformal transformation.

As discussed in the case of the conformal symmetry in the $U(1)$ gauge theory, this problem can be solved by considering the combined transformation, 
\begin{equation}
   \dl^\T_\zeta = \dl_\zeta -\dl_{\kappa_\zeta},
\end{equation}
where the gauge transformation $\dl_{\kappa_\zeta}$ is defined by (\ref{traceless mode gauge transformation}) with the mode-dependent gauge parameter $\kappa_\zeta^\mu$ which is given by $\kappa_\S^\mu$ in Appendix for the special conformal transformation generated by $Q_M^h$ and vanishes for the time translation and the rotation. These transformations form the closed algebra of the conformal symmetry such as
\begin{eqnarray}
     \dl^\T_\zeta h^{\T\T}_{ij} &=& i[Q^h_\zeta, h^{\T\T}_{ij}],
       \nonumber \\
     \dl^\T_\zeta h^\T_i &=& i[Q^h_\zeta, h^\T_i ],
       \nonumber \\
     \dl^\T_\zeta h &=& 0 ,
        \label{combined transformation for traceless mode}
\end{eqnarray}
which preserve the radiation$^+$ gauge.

\section{Physical States and Scaling Dimensions}
\setcounter{equation}{0}
\noindent

In a scale-invariant space-time, we can not have the ordinary particle picture such as propagating on a classical space-time any longer. Physical states are generated by conformal symmetry, and they are classified by the representation of conformal algebra \cite{hh, hamada05}, as in the case of two-dimensional quantum gravity states defined by the Virasoro conditions \cite{lz,bmp}. In this section, we examine such a four-dimensional quantum gravity state and its physical properties.

A conformally invariant vacuum annihilated by all the generators is uniquely determined by 
\begin{equation}
    | \Om \rangle = e^{-2b_1 \phi_0} |0\rangle ,
\end{equation}
where $\phi_0=\hat{q}/\sq{2b_1}$ is the zero mode of the conformal-mode field and $|0 \rangle$ is the standard Fock vacuum with the zero eigenvalue of $\hat{p}$. The exponential factor indicates the background charge coming from the $\hE_4 \phi$ term in the Wess-Zumino action. The physical states are spanned by the Fock space generated on the conformally invariant vacuum as
\begin{equation}
     |{\rm phys} \rangle = {\cal O}(a^\dag_{JM},b^\dag_{JM}, \cdots) |\Om \rangle .
\end{equation}
They satisfy the conformal invariance conditions \cite{hh,hamada05}
\begin{eqnarray}
         Q_M |{\rm phys} \rangle &=& 0,
             \nonumber  \\
         ( H-4)|{\rm phys} \rangle &=& R_{MN} |{\rm phys} \rangle =0 .
             \label{physical state condition}
\end{eqnarray}
The eigenvalue $4$ of the Hamiltonian indicates that the physical states has the conformal weight $4$, so that its volume integral has the vanishing weight in four dimensions. The ghost fields in the radiation$^+$ gauge, which have 15 degrees of freedom, are decoupled and considered to be integrated out. If we consider the full generators including the ghost sector, the Hamiltonian condition has the Wheeler-DeWitt form of $H=0$ \cite{amm97,hamada05}.

The physical state is now decomposed into four sectors: the scalar field, the gauge field, the conformal mode and the traceless tensor mode. Each sector consists of the Hamiltonian eigenstates satisfying the $Q_M$ condition. We first construct such states, and then impose the Hamiltonian and the rotation invariance conditions after combining all sectors. As examples, we here give the results for the scalar field and the conformal-mode field sectors.

In order to find states satisfying the $Q_M$ condition, we seek a creation operator that commutes with the generator. We first consider the scalar field sector. The commutator between the generator $Q_M^X$ and the creation mode is given by
\begin{equation}
    [ Q^X_M , \vphi^\dag_{J M_1} ] = \hbox{$\sq{2J(2J+1)}$} \sum_{M_2}
                    \eps_{M_2} \C^{\half M}_{J M_1, J-\half -M_2} \vphi^\dag_{J-\half M_2},
\end{equation}
and thus the creation operator that commutes with $Q^X_M$ is only the lowest mode $\vphi_{00}^\dag$ with the conformal weight $1$. We here impose the $Z_2$ symmetry $X \leftrightarrow -X$, and thus the odd products of the scalar field modes are removed below.

Next, we look for creation operators constructed in a bilinear form. Consider the operator with the conformal weight $2L+2$ belonging to the $(J,J)$ representation of the isometry group $SU(2)\times SU(2)$, denoted as
\begin{equation}
   \Phi^{[L]\dag}_{J N} = \sum^L_{K=0} \sum_{M_1}\sum_{M_2} f(L,K)\C^{JN}_{L-K M_1, K M_2}
                      \vphi^\dag_{L-K M_1} \vphi^\dag_{K M_2} .
\end{equation}
The commutator between the generator and this operator is computed as
\begin{eqnarray}
  && [ Q^X_M , \Phi^{[L]\dag}_{J N} ] 
     = \sum_{K=0}^L \sum_{M_1}\sum_{M_2} 
      \vphi^\dag_{L-K-\half M_1}\vphi^\dag_{K M_2} 
               \nonumber \\ 
  && \times 
      \sum_S  \biggl\{  \sq{(2L-2K)(2L-2K+1)} f(L,K)
                 \eps_S \C^{\half M}_{L-K-\half M_1, L-K -S} 
                            \C^{J N}_{L-K S, K M_2} 
              \nonumber \\ 
  && \quad
         +  \sq{(2K+1)(2K+2)} f \left( L, K+\half \right)
               \eps_S \C^{\half M}_{K M_2, K+\half -S} 
          \C^{J N}_{K+\half S, L-K-\half M_1}  \biggr\} .
                \nonumber \\
\end{eqnarray}
Using the crossing properties of the $SU(2)\times SU(2)$ Clebsch-Gordan coefficients (\ref{crossing relation}), we find that the right-hand side vanishes if and only if $J=L$ and $L$ is a positive integer, and the function $f$ satisfies the equation
\begin{equation}
    f \left( L,K+\half \right) = - \sq{\fr{(2L-2K)(2L-2K+1)}{(2K+1)(2K+2)}} f(L,K) .
\end{equation}
Solving this recursion relation, we obtain
\begin{equation}
     f(L,K)= \fr{(-1)^{2K}}{\sq{(2L-2K+1)(2K+1)}} 
                   \left( \begin{array}{c}
                                     2L \\
                                     2K 
                                     \end{array} \right) 
\end{equation}
up to the $L$-dependent normalization. Thus, we obtain the $Q^X_M$ invariant creation operators, denoted as $\Phi^\dag_{L N} = \Phi^{[L]\dag}_{L N}$ below.

By joining these operators using the $SU(2)\times SU(2)$ Clebsch-Gordan coefficients, we can construct the basis of $Q_M$-invariant creation operators in the scalar field sector. Due to the crossing properties of the Clebsch-Gordan coefficients, any $Q_M$-invariant creation operators will be expressed in such a fundamental form. Thus, this operator is expected to be the building block of physical states in the scalar field sector.

Similarly, we can construct building blocks for the conformal-mode field sector. The commutators between the $Q^\phi_M$ generator and the zero modes are given by
\begin{eqnarray}
     \left[ Q^\phi_M, \hat{q} \right] &=& -a_{\half M}, 
          \nonumber  \\
     \left[ Q^\phi_M, \hat{p} \right] &=& 0 .   
\end{eqnarray}
For the creation mode $a^\dag$, we obtain
\begin{eqnarray}
    \left[ Q^\phi_M, a^\dag_{\half M_1} \right]
    &=& ( \hbox{$\sq{2b_1}$}-i\hat{p} ) \dl_{M,M_1},
             \nonumber  \\
   \left[ Q^\phi_M, a^\dag_{J M_1} \right]
   &=& \a \left( J-\half \right) \sum_{M_2} 
     \eps_{M_2} \C^{\half M}_{J M_1, J-\half -M_2}  a^\dag_{J-\half M_2},
\end{eqnarray}
where $J \geq 1$. For $b^\dag$, we obtain
\begin{eqnarray}
   \left[ Q^\phi_M, b^\dag_{J M_1} \right]
   &=& -\gm (J) \sum_{M_2} \eps_{M_2} \C^{\half M}_{J M_1, J+\half -M_2} a^\dag_{J+\half M_2}
           \nonumber \\
   &&  -\b \left( J-\half \right) \sum_{M_2} 
        \eps_{M_2} \C^{\half M}_{J M_1, J-\half -M_2} b^\dag_{J-\half M_2},
\end{eqnarray}
where $J \geq 0$.

Since there is no creation mode that commutes with $Q^\phi_M$, we look for operators constructed in a bilinear form, as in the case of the scalar field sector. Using the crossing properties of the $SU(2)\times SU(2)$ Clebsch-Gordan coefficients, we find two $Q^\phi_M$ invariant combinations with the conformal weight $2L$:
\begin{eqnarray}
    S^{\dag}_{L N}
     &=& \chi(\hat{p}) a^{\dag}_{L N}
         + \sum_{K=\half}^{L-\half} \sum_{M_1}\sum_{M_2} x(L,K)
         \C^{L N}_{L-K M_1, K M_2} a^{\dag}_{L-K M_1} a^{\dag}_{K M_2},
                \nonumber \\
    {\cal S}^{\dag}_{L-1 N}
     &=& \psi(\hat{p}) b^{\dag}_{L-1 N}
         + \sum_{K=\half}^{L-\half} \sum_{M_1}\sum_{M_2} x(L,K)
         \C^{L-1 N}_{L-K M_1, K M_2} a^{\dag}_{L-K M_1} a^{\dag}_{K M_2}
            \nonumber  \\
      && + \sum_{K=\half}^{L-1} \sum_{M_1,M_2} y(L,K)
        \C^{L-1 N}_{L-K-1 M_1, K M_2}
        b^{\dag}_{L-K-1 M_1} a^{\dag}_{K M_2}
\end{eqnarray}
for integers $L \geq 1$. The coefficients are given by
\begin{eqnarray}
      x(L,K) &=& \fr{(-1)^{2K}}{\sq{(2L-2K+1)(2K+1)}}\sq{ \left( \begin{array}{c}
                                     2L \\
                                     2K
                                     \end{array} \right)
                             \left(   \begin{array}{c}
                                     2L-2 \\
                                     2K-1
                                     \end{array} \right) },
              \nonumber \\
      y(L,K) &=& -2\sq{(2L-2K-1)(2L-2K+1)}x(L,K) .
\end{eqnarray}
For any half-integer $L$, these functions vanish. The $\hat{p}$-dependent operators are given by
\begin{eqnarray}
      \chi (\hat{p}) &=& \fr{1}{\sq{2(2L-1)(2L+1)}}(\hbox{$\sq{2b_1}$}-i\hat{p}),
                 \nonumber \\
      \psi (\hat{p}) &=& -\sq{2} ( \hbox{$\sq{2b_1}$}-i\hat{p})  .        
\end{eqnarray}
These two types of operators are expected to be the building blocks of physical states in the conformal-mode field sector, which are summarized in Table \ref{building block for conformal mode}.

\begin{table}[h]
\begin{center}
\begin{tabular}{|c|c|}  \hline
rank of tensor index & $0$  \\ \hline
creation operator          & $S^\dag_{LN}$  \\
                      & ${\cal S}^\dag_{L-1N}  $  \\
weight $(L \in {\bf Z}_{\geq 1})$ &  $2L$    \\ \hline
\end{tabular}
\end{center}
\caption{\label{building block for conformal mode}Building blocks in the conformal-mode field sector}
\end{table}

The creation mode that commutes with $Q^h_M$ in the traceless tensor field sector is only the lowest positive-metric mode, $c^\dag_{1(Mx)}$, in the transverse-traceless field $h^{\T\T}_{ij}$. The $Q^h_M$-invariant creation operators constructed as a bilinear form of the creation modes are rather complicated. We have to consider such operators including the tensor indices up to rank $4$. We do not here present the explicit forms of such operators, which have been classified in \cite{hamada05} using generalized crossing properties for tensor harmonics. They will give the building blocks for the traceless tensor field sector listed in Table \ref{building block for traceless mode}. Any $Q^h_M$-invariant states are expected to be constructed from these building blocks.

\begin{table}[h]
\begin{center}
\begin{tabular}{|c|ccccc|}  \hline
rank of tensor index & $0$  & $1$ & $2$ &$3$ & $4$    \\ \hline
creation operator  & $A^\dag_{LN}$ & $B^\dag_{L-\half(Ny)}$ & $c^\dag_{1 (Nx)}$
              & $D^\dag_{L-\half (Nz)}$ & $E^\dag_{L(Nw)}$  \\
              & ${\cal A}^\dag_{L-1 N}$ &  &  &  & ${\cal E}^\dag_{L-1 (Nw)}$  \\
weight $(L \in {\bf Z}_{\geq 3})$ &  $2L$  & $2L$  & $2$ & $2L$ & $2L$       \\ \hline
\end{tabular}
\end{center}
\caption{\label{building block for traceless mode} Building blocks in the traceless tensor field sector}
\end{table}

We now construct the physical state satisfying all the conditions (\ref{physical state condition}). Firstly, consider the states that depends only on the zero mode of the conformal-mode field satisfying the $Q_M$-invariance condition, which is given by
\begin{equation}
     |p,\Om \rangle = e^{ip\hat{q}}|\Om \rangle = e^{ip\sq{2b_1}\phi_0} |\Om \rangle .
\end{equation}
This is the eigenstate of $\hat{p}$ with the eigenvalue $p+i\sq{2b_1}$. The general state satisfying the conditions for $Q_M$ and $R_{MN}$ is constructed by acting those building blocks on the state $|p,\Om \rangle$, with all the tensor indices contracted by using the $SU(2)\times SU(2)$ Clebsch-Gordan coefficients. This state is denoted by ${\cal R}_n(S^\dag,\cdots)|p,\Om \rangle$, where the operator ${\cal R}_n$ carries the conformal weight $n$ of an even integer. The Hamiltonian condition gives the equation $(p+i\sq{2b_1})^2/2 + b_1+n=4$, so that $p$ should have the purely imaginary value $-i\gm_n/\sq{2b_1}$ with\footnote{ 
The Wess-Zumino coefficient (\ref{Wess-Zumino coefficient}) satisfies $b_1 > 4$ for non-negative numbers, $N_X$, $N_W$ and $N_A$.
}  
\begin{eqnarray}
     \gm_n &=& 2b_1 \left( 1- \sq{1-\fr{4-n}{b_1}} \right)
              \nonumber \\
     &=& 4-n + \fr{1}{4b_1}(4-n)^2 + o( 1/b_1^2 ),
\end{eqnarray}
and thus we obtain the physical state
\begin{equation}
      {\cal R}_n (S^\dag, \cdots) e^{\gm_n \phi_0} |\Om \rangle .
\end{equation}
Here, the solution that $\gm_n$ approaches the canonical value $4-n$ in the large $b_1$ limit is selected. For each gravitational state, there is a field operator ${\cal O}$ such that the state is given by the limit: $|{\rm phys}\rangle = \lim_{\eta \to i\infty} e^{-i4\eta}{\cal O} (\eta,\bx)|\Om \rangle $.

As examples, we show the lower $n$ gravitational states up to $4$ coupled to the scalar field in the followings. The lowest weight state is the gravitationally dressed state of the identity operator,
\begin{equation}
    e^{\gm_0 \phi_0} |\Om \rangle ,
\end{equation}
which corresponds to the cosmological constant, or the physical metric field, $\sq{-g}$. For $n=2$, there are two gravitational states,
\begin{equation}
    {\cal S}^\dag_{00} e^{\gm_2 \phi_0}|\Om \rangle, \qquad
    \Phi^\dag_{00}e^{\gm_2 \phi_0}|\Om \rangle .
\end{equation}
The left-hand side corresponds to the scalar curvature, $\sq{-g}R$, and the right-hand side is the gravitationally dressed scalar field, $\sq{-g}X^2$. For $n=4$, there are five gravitational states,
\begin{eqnarray}
    && \sum_{N,x} \eps_N c^\dag_{1(-Nx)}c^\dag_{1(Nx)} |\Om \rangle,  \quad
      ({\cal S}^\dag_{00} )^2 |\Om \rangle,  \quad
     \sum_{N} \eps_N S^\dag_{1-N}S^\dag_{1 N} |\Om \rangle, 
         \nonumber \\
    && 
     \Phi^\dag_{00} {\cal S}^\dag_{00} |\Om \rangle,  \quad
     ( \Phi^\dag_{00})^2 |\Om \rangle ,
\end{eqnarray}
where $\gm_4=0$ is taken into account. The first state corresponds to the square of the Weyl tensor, $\sq{-g}C^2_{\mu\nu\lam\s}$, and the second is the square of the scalar curvature, $\sq{-g}R^2$. The third is a diffeomorphism invariant state independent of the first two states. The last two states are the dressed scalar fields, $\sq{-g}R X^2$ and $\sq{-g}X^4$, respectively.

The gravitational corrections has the purely imaginary value of the zero-mode momentum $p$.  This is a peculiar property of quantum diffeomorphism invariant states. If the zero-mode momentum were real, the conformal field could be normalizable in the sense of delta function as $\int d\phi_0 e^{ip^\pp\phi_0}e^{-ip \phi_0}=\dl(p^\pp -p)$. The pure imaginary value implies that the diffeomorphism invariant state is real, and thus not normalizable in the usual sense. In order to evaluate two point correlation functions, we have to introduce the potential term with the zero-mode charge  $\gm_n$ such as the Einstein term to settle the zero-mode integral, as in the case of two-dimensional quantum gravity \cite{seiberg, gl, hamada94}. The correlation function has a power-law behavior with respect to the mass scale in the potential term.

Although it is difficult to calculate such a two-point correlation function, we can evaluate the scaling dimension of the physical conformal field from the scale transformation property. Consider that the conformal field ${\cal O}_n$ with the zero-mode charge $\gm_n$ has the scaling dimension $\Delta_n$, and it transforms as
\begin{equation}
    d^4 x {\cal O}_n \to \om^{4-\Delta_n} d^4 x {\cal O}_n
\end{equation}
under the constant Weyl rescaling defined such that the cosmological constant field transforms as $\Delta_0=0$. The Weyl rescaling is equivalent to the constant shift of the zero-mode, $\phi_0 \to \phi_0+(4/\gm_0)\ln \om$. By this shift, the conformal field $d^4 x {\cal O}_n$ changes to $\om^{4\gm_n/\gm_0}d^4 x{\cal O}_n$. Thus, we obtain the relation  
\begin{equation}
    \Delta_n = 4 - 4 \fr{\gm_n}{\gm_0} 
       \label{4 dim. conformal dimension}
\end{equation}
for even integers $n >0$. This is the physical scaling dimension of the conformal field ${\cal O}_n$ satisfying the bound of $\Delta_n >1$ \cite{mack, kluwer}, which approaches the canonical value $n$ at the large $b_1$ limit. The scaling behavior of the two-point correlation of ${\cal O}_n$ is determined by the scaling dimension $\Delta_n$.

These diffeomorphism invariant physical states are composite states in which the positive-metric and the negative-metric modes are mixed due to conformal symmetry and the negative-metric mode does not appear independently at all. This suggests that the correctness of the overall sign of the gravitational action given in a diffeomorphism invariant combination, not the sign of each mode, is significant for unitarity. Since the gravitational actions are bounded below as discussed in section 2 and the diffeomorphism invariance seems to force the physical state to be real quantum mechanically, it is expected that the amplitude of their two-point correlation function becomes positive due to no factor violating the reality in the viewpoint of symmetry.

\section{Conclusion and Discussion}
\setcounter{equation}{0}
\noindent

In this paper, we examined quantum diffeomorphism symmetry in four-dimensional quantum gravity on the cylindrical background $R \times S^3$. We showed that conformal symmetry is equal to a residual gauge symmetry of diffeomorphism invariance in the radiation$^+$ gauge. We also showed that the conformal transformation preserving the gauge-fixing condition that forms a closed algebra is given by a combination of naive conformal transformation and gauge transformation with a certain mode-dependent parameter.

The conformal invariance forces us change the aspect of space-time at high energies above the Planck scale, where a traditional S-matrix description is not adequate at all. Consequently, this requires a new prescription to deal with negative-metric modes which can be carried out by making use of the conformal symmetry.

The physical state in such a non-perturbative regime was constructed in terms of the composite conformal field by solving the conformal invariance condition, and its physical scaling dimension was calculated. Then, the unitarity issue of gravity was discussed in the context of conformal field theory. It is suggested that since the renormalizable gravitational action has the right sign ensuring that the path integral is well-defined, the diffeomorphism invariance seems to preserve the reallity of the conformal field like the scalar curvature quantum mechanically and thus it is expected that its two-point correlation function becomes positive. The two-point correlation of the scalar curvature will give a power-law spectrum of the universe in the initial stage of inflation \cite{hy, hhy, hhsy}.

We here give a brief comment on the unitarity argument \cite{tomboulis,ft,ft-report} done in 1970's based on the idea of Lee and Wick \cite{lw, nakanishi}. The essence of their idea is that the positive-metric and the negative-metric modes in a higher-derivative field are mixed by interactions so that the ghost pole in the resummed propagator disappears from the real axis due to radiative corrections in the case of asymptotically free theory. Although this idea is still meaningful when we discuss the connection with real world, one can not avoid the appearance of the asymptotic ghost state after all, because in those days higher-derivative models have no symmetry mixing the positive-metric and the negative-metric modes so that the ghost mode becomes gauge invariant when the interaction turns off at the vanishing coupling limit. Furthermore, the asymptotic freedom means that the perturbative picture that free particles propagate in Minkowski space-time arises at very high energies. On the other hand, in our model it implies that there is no classical space-time to define such particle states, but totally fluctuating quantum space-time with exact conformal symmetry, which mixes gravitational modes in a diffeomorphism invariant manner.

The asymptotic state should be defined by the classical limit $\hbar \to 0$. As discussed in section 2, since $\hbar$ appears in front of the lower-derivative action, four-derivative gravitational fields describe purely quantum mechanical virtual states. Thus, the asymptotic state exists only at low energies below the dynamical energy scale $\Lam_\QG$ where the dynamics is ruled by the Einstein action. If we wish to define the S-matrix, we have to prepare the asymptotic state far from a place where quantum gravity turns on such as the center of a black hole.

Finally, we give a comment on another formulation of quantum gravity: a four-dimensional simplicial quantum gravity based on the dynamical triangulation. This is a formulation adopting the background-metric independence as the first principle. The path integral over metric function is replaced by the summation over all space-time configurations numerically in a simplicial manifold. In the recent analysis it has been recognized that these two methods belong to the same universality class \cite{hey, nova}.

\vspace{1cm}

\appendix

\begin{center}
{\Large {\bf Appendix}}
\end{center}

\section{Gauge Parameter $\kappa^\mu_\S$}
\setcounter{equation}{0}
\noindent

The parameter $\kappa_\S^\mu=(\kappa_\S^0,\kappa_\S^k)$ in the combined transformation for the traceless tensor field is given by
\begin{eqnarray}
   (\kappa^0_\S)_M &=& - \fr{3}{8} \sum_{J \geq 1} \fr{1}{\sq{(2J-1)(2J+1)(2J+3)}} 
              \nonumber \\
      && \quad \times 
        \biggl\{ \fr{1}{2J} \sum_{N,y}\sum_S \eps_S \G^{\half M}_{J(Ny);J-S}e_{J(Ny)}
                                          e^{-i2J\eta} Y_{JS} 
                        \nonumber \\
     && \qquad            
          + \fr{1}{2J+2} \sum_{N,y}\sum_S (-\eps_N) \G^{\half M}_{J(-Ny);JS}e^\dag_{J(Ny)}
                                          e^{i(2J+2)\eta} Y^*_{JS}  \biggr\} 
               \nonumber 
\end{eqnarray}
and
\begin{eqnarray}
   && (\kappa^k_\S)_M =
         \nonumber \\
      && \fr{i}{4} \sum_{J \geq 1} \fr{1}{\sq{(2J-1)(2J+1)(2J+3)}} 
               \nonumber \\
       && \times  
         \biggl\{  \fr{1}{2J} \sum_{N,y}\sum_{V,y^\pp}
             (-\eps_V) \D^{\half M}_{J(Ny),J+\half(-Vy^\pp)}
                      e_{J(Ny)} e^{-i2J\eta} Y^k_{J+\half(Vy^\pp)}
                   \nonumber \\
        && \quad
           -\fr{1}{2J+2} \sum_{N,y}\sum_{V,y^\pp}
               (-\eps_V) \D^{\half M}_{J(Ny),J-\half(-Vy^\pp)}
                     e_{J(Ny)} e^{-i2J\eta} Y^k_{J-\half(Vy^\pp)}
                   \nonumber \\
        && \quad
           - \fr{1}{2J} \sum_{N,y}\sum_{V,y^\pp}
                (-\eps_N) \D^{\half M}_{J(-Ny),J+\half(Vy^\pp)}
                      e^\dag_{J(Ny)} e^{i(2J+2)\eta} Y^{k*}_{J+\half(Vy^\pp)}
                   \nonumber \\
       && \quad
          + \fr{1}{2J+2} \sum_{N,y}\sum_{V,y^\pp}
               (-\eps_N) \D^{\half M}_{J(-Ny),J-\half(Vy^\pp)}
                      e^\dag_{J(Ny)} e^{i(2J+2)\eta} Y^{k*}_{J-\half(Vy^\pp)}
                \biggr\}
                  \nonumber \\
     && + \fr{i}{8} \sum_{J \geq 1} \fr{1}{\sq{J(2J+1)}} 
                  \nonumber \\
     && \times
        \biggr\{  -\fr{1}{2J-1} \sum_{N,x}\sum_{V,y^\pp} 
               (-\eps_V) \H^{\half M}_{J(Nx);J(-Vy^\pp)}
                     c_{J(Nx)} e^{-i(2J-1)\eta} Y^k_{J(Vy^\pp)}
                  \nonumber \\
     && \quad
         +\fr{2J-3}{(2J-1)(2J+3)} \sum_{N,x}\sum_{V,y^\pp} 
                 \eps_N \H^{\half M}_{J(-Nx);J(Vy^\pp)}
                     c^\dag_{J(Nx)} e^{i(2J+1)\eta} Y^{k*}_{J(Vy^\pp)}
            \biggr\}
                  \nonumber \\
    && + \fr{i}{8} \sum_{J \geq 1} \fr{1}{\sq{(J+1)(2J+1)}} 
                  \nonumber \\
    && \times \biggr\{ 
        -\fr{2J+5}{(2J-1)(2J+3)} \sum_{N,x}\sum_{V,y^\pp} 
              (-\eps_V) \H^{\half M}_{J(Nx);J(-Vy^\pp)}
                     d_{J(Nx)} e^{-i(2J+1)\eta} Y^k_{J(Vy^\pp)}
                  \nonumber \\
    && \quad
         +\fr{1}{2J+3} \sum_{N,x}\sum_{V,y^\pp} 
                \eps_N \H^{\half M}_{J(-Nx);J(Vy^\pp)}
                     d^\dag_{J(Nx)} e^{i(2J+3)\eta} Y^{k*}_{J(Vy^\pp)}
            \biggr\}
                  \nonumber \\
    && +\fr{i}{8} \sum_{J \geq 1} \fr{1}{\sq{(2J-1)(2J+1)(2J+3)}}
                  \nonumber \\
    && \times 
        \fr{1}{2J(2J+2)} \biggl\{
          - \sum_{N,y}\sum_S \eps_S \G^{\half M}_{J(Ny);J-S} 
                e_{J(Ny)} e^{-i2J\eta} \hnb^k Y_{JS}
                 \nonumber \\
    && \qquad\qquad\qquad
         + \sum_{N,y}\sum_S (-\eps_N) \G^{\half M}_{J(-Ny);JS} 
          e^\dag_{J(Ny)} e^{i(2J+2)\eta} \hnb^k Y^*_{JS}
       \biggr\}.
                \nonumber
\end{eqnarray}

The product expansions used to determine $\kappa^\mu_\S$ are 
\begin{eqnarray}
    && Y_{\half M}^* Y^{ij}_{J(Nx)}
          \nonumber \\
    && = \fr{1}{\sq{\V3}}
         \biggl\{ \sum_{T,x^\pp} \E^{\half M}_{J(Nx),J+\half (Tx^\pp)}Y^{ij*}_{J+\half (Tx^\pp)}
               + \sum_{T,x^\pp} \E^{\half M}_{J(Nx),J-\half (Tx^\pp)}Y^{ij*}_{J-\half (Tx^\pp)}
             \nonumber \\
     && \qquad\qquad
               +  \fr{2}{(2J-1)(2J+3)} \sum_{V,y^\pp} \H^{\half M}_{J(Nx);J(Vy^\pp)} 
                \hnb^{(i} Y^{j)*}_{J(Vy^\pp)} 
               \biggr\},
                   \nonumber \\
   && \hnb^k Y_{\half M}^* \hnb_k Y^{ij}_{J(Nx)} 
          \nonumber \\
   && = \fr{1}{\sq{\V3}}
       \biggl\{ -2J\sum_{T,x^\pp} \E^{\half M}_{J(Nx),J+\half (Tx^\pp)}Y^{ij*}_{J+\half (Tx^\pp)}
             \nonumber \\
   && \qquad\qquad
          +(2J+2) \sum_{T,x^\pp} \E^{\half M}_{J(Nx),J-\half (Tx^\pp)}Y^{ij*}_{J-\half (Tx^\pp)} 
             \nonumber \\
    && \qquad\qquad
          +  \fr{6}{(2J-1)(2J+3)} \sum_{V,y^\pp} \H^{\half M}_{J(Nx);J(Vy^\pp)} 
            \hnb^{(i} Y^{j)*}_{J(Vy^\pp)}
          \biggr\} ,  
            \nonumber \\
    && \hnb^{(i}Y^*_{\half M} Y^{j)}_{J(Ny)} 
       -\fr{1}{3}\hgm^{ij} \hnb_k \left( Y^*_{\half M} Y^k_{J(Ny)} \right)
            \nonumber \\
    && = \fr{1}{\sq{\V3}}
         \biggl\{ - \sum_{T,x^\pp} \H^{\half M}_{J(Tx^\pp);J(Ny)} Y^{ij*}_{J(Tx^\pp)}
            \nonumber \\
    && \qquad\qquad
          + \fr{1}{2J} \sum_{V,y^\pp} \D^{\half M}_{J(Ny),J+\half (Vy^\pp)} 
            \hnb^{(i}Y^{j)*}_{J+\half (Vy^\pp)}
              \nonumber \\
    && \qquad\qquad
          -\fr{1}{2J+2} \sum_{V,y^\pp} \D^{\half M}_{J(Ny),J-\half (Vy^\pp)}
            \hnb^{(i}Y^{j)*}_{J-\half (Vy^\pp)}
              \nonumber \\
    && \qquad\qquad
       - \half \fr{1}{2J(2J+2)}\sum_S \G^{\half M}_{J(Ny);JS} 
          \left( \hnb^i \hnb^j - \fr{1}{3}\gm^{ij} \hnb^2 \right) Y^*_{JS} 
        \biggr\} ,
             \nonumber
\end{eqnarray}
where the symmetric product is denoted as $a^{(i}b^{j)}=(a^i b^j +a^j b^i)/2$.


\end{document}